\documentclass[utf8,arxiv]{frontiersSCNS} 

\usepackage{url,hyperref,lineno,microtype,subcaption}
\hypersetup{
    colorlinks=true,
    linkcolor=cyan,
    filecolor=gray,      
    urlcolor=blue,
    citecolor=darkgray
}
\usepackage[onehalfspacing]{setspace}
\usepackage{amsmath}
\usepackage{xspace}
\usepackage{multirow}
\usepackage{algorithm}
\usepackage{algpseudocode}
\usepackage{ulem}

\ifarxiv
\else
\linenumbers
\fi

\def\keyFont{\fontsize{8}{11}\helveticabold }
\def\firstAuthorLast{Elabd {et~al.}} 
\def\Authors{
Abdelrahman Elabd\,$^{1}$,
Vesal Razavimaleki\,$^{2}$,
Shi-Yu Huang\,$^{3}$,
Javier Duarte\,$^{2}$,
Markus Atkinson\,$^{4}$,
Gage DeZoort\,$^{5}$,
Peter Elmer\,$^{5}$,
Scott Hauck\,$^{6}$,
Jin-Xuan Hu\,$^{3}$,
Shih-Chieh Hsu\,$^{6,7}$,
Bo-Cheng Lai\,$^{3}$,
Mark Neubauer\,$^{4}$,
Isobel Ojalvo\,$^{5}$,
Savannah Thais\,$^{5}$ and
Matthew Trahms\,$^{6}$
  }
\def\Address{
$^{1}$Department of Physics \& Astronomy, University of Pennsylvania, Philadelphia, PA, United States, 
$^{2}$Department of Physics, University of California San Diego, La Jolla, CA, United States,
$^{3}$Department of Electrical \& Computer Engineering, National Yang Ming Chiao Tung University, Taipei, Taiwan,
$^{4}$Department of Physics, University of Illinois at Urbana-Champaign, Champaign, IL, United States,
$^{5}$Department of Physics, Princeton University, Princeton, NJ, United States,
$^{6}$Department of Electrical \& Computer Engineering, University of Washington, Seattle, WA, United States,
$^{7}$Department of Physics, University of Washington, Seattle, WA, United States
}

\def\corrAuthor{Mark Neubauer}
\def\corrEmail{msn@illinois.edu}

\newcommand{\apfixed}[2]{\texttt{ap\_fixed<#1,#2>}\xspace}
\newcommand{\pt}{\ensuremath{p_{\mathrm{T}}}\xspace}
\newcommand{\ptmin}{\ensuremath{\pt^\mathrm{min}}\xspace}
\newcommand{\GeV}{\ensuremath{\,\text{Ge\hspace{-.08em}V}}\xspace}

\newcommand{\unit}[1]{\,\text{#1}\xspace}
\newcommand{\hlsfml}{\texttt{hls4ml}\xspace}

\newcommand{\brevitas}{\textsc{Brevitas}\xspace}

\newcommand{\nnodes}{\ensuremath{n_\mathrm{nodes}\xspace}}
\newcommand{\nedges}{\ensuremath{n_\mathrm{edges}\xspace}}

\begin{document}

\title[GNNs for Tracking on FPGAs]{Graph Neural Networks for Charged Particle Tracking on FPGAs}

\onecolumn
\firstpage{1}

\author[\firstAuthorLast ]{\Authors} 
\address{} 
\correspondance{} 

\extraAuth{Javier Duarte\\
jduarte@ucsd.edu}

\maketitle

\begin{abstract}
\section{}
The determination of charged particle trajectories in collisions at the CERN Large Hadron Collider (LHC) is an important but challenging problem, especially in the high interaction density conditions expected during the future high-luminosity phase of the LHC (HL-LHC). 
Graph neural networks (GNNs) are a type of geometric deep learning algorithm that has successfully been applied to this task by embedding tracker data as a graph---nodes represent hits, while edges represent possible track segments---and classifying the edges as true or fake track segments. 
However, their study in hardware- or software-based trigger applications has been limited due to their large computational cost.
In this paper, we introduce an automated translation workflow, integrated into a broader tool called \hlsfml, for converting GNNs into firmware for field-programmable gate arrays (FPGAs).
We use this translation tool to implement GNNs for charged particle tracking, trained using the TrackML challenge dataset, on FPGAs with designs targeting different graph sizes, task complexites, and latency/throughput requirements.
This work could enable the inclusion of charged particle tracking GNNs at the trigger level for HL-LHC experiments.
\tiny
 \keyFont{\section{Keywords:} graph neural networks, FPGAs, tracking, LHC, trigger} 
\end{abstract}

\section{Introduction}

In high energy physics (HEP), charged particle tracking~\citep{Amrouche:2019wmx,Strandlie:2010zz} is a crucial task necessary for the accurate determination of the kinematics of the particles produced in a collision event, including the position, direction, and momentum of the particles at their production points. 
This task leverages specialized detectors positioned close to the beam collision area in a strong magnetic field.
When charged particles are created in the collisions, their trajectories bend in the magnetic field and they ionize the material of these detectors as they move outwards from the production point, providing position measurements along the trajectory of each particle. 
The objective of tracking algorithms is to identify the individual trajectories of these charged particles and extract relevant particle kinematics.
Current tracking algorithms~\citep{Chatrchyan:2014fea,Aaboud:2017all,combkalman1,combkalman2,combkalman3,kalman} scale worse than quadratically with the number of hits, which is expected to increase dramatically at higher beam intensities due to the presence of simultaneous proton-proton interactions (or \textit{pileup}) and for more granular, more sensitive detectors.
This motivates the study of alternative algorithms with different computational scaling. 
Another important consideration is the ability to accelerate these algorithms using highly-parallel heterogeneous computing resources like graphics processing units (GPUs) and field-programmable gate arrays (FPGAs) as further improvements in single-core CPU performance may be limited~\citep{breakdown,dennard}.
Recent efforts~\citep{Farrell:2018cjr,Ju:2020xty,Dezoort:2021kfk} have demonstrated the effectiveness of graph neural networks (GNNs) to correctly classify ``segments'' belonging to tracks.
Graph-based approaches are well suited to this task because tracking data can be naturally encoded as a graph structure~\citep{Shlomi:2020gdn} and GNNs consider local information between pairs of hits to learn relationships between them in order to ``connect the dots'' and infer tracks.
In other words, track finding is an example of edge classification on a graph data structure.

In this paper, we develop an automatic translation toolflow to convert GNNs~\citep{Dezoort:2021kfk} for segment classification, based on the interaction network (IN) architecture~\citep{Battaglia:2016jem,battaglia2018relational}, to FPGA firmware.
FPGA implementations enable efficient processing, in both speed and energy consumption for large HEP datasets.
They may also enable the use of GNNs in the high-throughput, FPGA-based data filter system, known as the level-1 trigger (L1T)~\citep{ATLASL1T,ATLASP2L1T,CMSL1T,CMSP2L1T}, which has strict sub-microsecond latency requirements that only FPGAs or application-specific integrated circuits (ASICs) can meet.
For instance, in the upgraded CMS design, a latency of 4\,$\mu$s (plus 1\,$\mu$s for data transfer) and an event throughput of 2.22\unit{MHz} are required for the L1 track trigger~\citep{CMSP2L1T}.
Conversely, in situations when a longer latency is permissible such as the high-level trigger (HLT)~\citep{Trocino:2014jya} or offline processing, they may be used in coprocessing applications and scale to larger tasks.
Our automatic translation code is integrated with \hlsfml~\citep{Duarte:2018ite,vloncar_2021_5680908}, a more general compiler for converting machine learning (ML) algorithms into FPGA firmware.
We evaluate the resource usage, latency, and tracking performance of a variety of different implementations based on the benchmark TrackML dataset~\citep{Amrouche:2019wmx}.

This paper is structured as follows. 
In Section~\ref{sec:related}, we briefly recapitulate related work.
Section~\ref{sec:data} defines the benchmark TrackML dataset and task, including the data preprocessing and graph encoding.
Section~\ref{sec:IN} describes the IN model used for the track segment classification task.
Section~\ref{sec:hls4ml} describes the \hlsfml user interface, while section~\ref{sec:FPGA} describes the FPGA designs.
Section~\ref{sec:results} summarizes the results of the FPGA firmware synthesis, including measurements of performance, timing, and FPGA resources.
Finally, a summary and outlook are given in Section~\ref{sec:summary}.

\section{Related work}
\label{sec:related}
GNNs have been explored for particle physics applications~\citep{Shlomi:2020gdn,Duarte:2020ngm}, including jet identification~\citep{Qu:2019gqs,Moreno:2019bmu,Moreno:2019neq}, pileup mitigation~\citep{ArjonaMartinez:2018eah,li2021semisupervised}, calorimeter energy measurements~\citep{Qasim:2019otl}, particle-flow reconstruction~\citep{Kieseler:2020wcq,Pata:2021oez}, and charged particle tracking~\citep{Farrell:2018cjr,Dezoort:2021kfk,Ju:2021ayy}.
Automatic translation of ML algorithms into FPGA firmware has also been studied for HEP tasks.
Using \hlsfml, several implementations for HEP-specific tasks have been provided for fully connected neural networks (NNs), autoencoders, boosted decision trees, and convolutional NNs~\citep{Duarte:2018ite,Summers:2020xiy,DiGuglielmo:2020eqx,Coelho:2020zfu,Aarrestad:2021zos,Govorkova:2021utb}.
This tool has also been applied extensively for tasks in the HL-LHC upgrade of the CMS L1T system, including anomaly detection, muon energy regression and identification, tau lepton identification, and vector boson fusion event classification~\citep{CMSP2L1T}. 
Moreover, a GNN model known as a GarNet was studied for calorimeter energy regression and deployed on FPGAs using \hlsfml in \cite{Iiyama:2020wap}.
Our current work extends this by allowing more generic IN architectures to be converted with \hlsfml, permitting the specification of a variable graph adjacency matrix as input, and supporting per-node or per-edge outputs as well as per-graph outputs.
This paper also supersedes an earlier version of this work in \cite{Heintz:2020soy}.

Hardware acceleration of GNN inference, and graph processing in general, has been an active area of study~\citep{besta2019graph,gui2019survey,9361829}.
\cite{6861577,10.1145/3007787.3001155,10.1145/3373087.3375312, yan2020hygcn,9218751,geng2020awbgcn,kiningham2020grip} describe various other examples of GNN acceleration architectures. 
While these frameworks are applicable to various graph processing tasks, they  may not apply to the strict latency requirements of the LHC trigger and they typically require the user to specify the design in a highly specialized format.

\section{TrackML data}
\label{sec:data}
The results presented in this paper are based on the TrackML dataset. 
The TrackML dataset is a simulated set of proton-proton collision events originally developed for the TrackML Particle Tracking Challenge~\citep{Amrouche:2019wmx}. 
Each event is generated with 200 pileup interactions on average, simulating the high-pileup conditions expected at the HL-LHC. 
Each event data record contains 3D hit positions, additional ``cell'' information (e.g. directional information), hit truth information, including the true location of each hit, and the features of the true charged particle that generated each hit. 
In particular, truth particles are specified by particle IDs ($p_\mathrm{ID}$) and three-momentum vectors ($\mathbf{p}$). 

\begin{figure}[t]
    \begin{center}
    \includegraphics[width=\columnwidth]{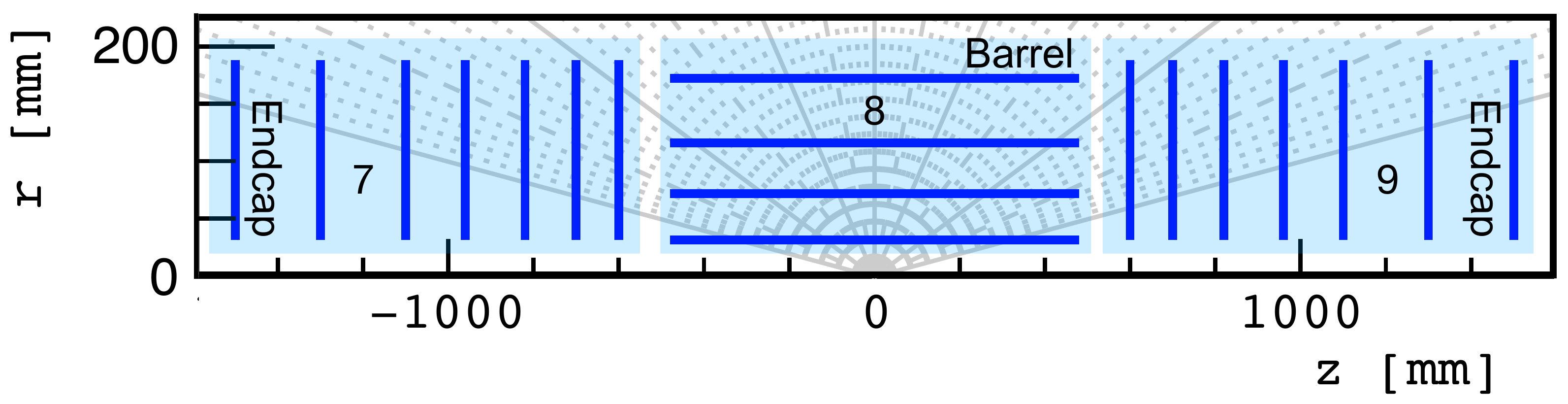}
    \end{center}
    \caption{
    TrackML pixel detector regions labeled 7, 8, and 9, used in this study, consisting of 7, 4, and 7 layers, respectively.}
    \label{fig:pixeldetector}
\end{figure}

The TrackML detector emulates a generic LHC silicon-based tracker, containing discrete layers of silicon sensors immersed in a strong, inhomogeneous magnetic field. 
We focus specifically on the innermost silicon pixel layers, a highly-granular set of 4 barrel and 14 endcap layers in the innermost tracker regions with a spatial resolution of 50\unit{$\mu$m}$\times$50\unit{$\mu$m}. 
These pixel layers are illustrated in Figure~\ref{fig:pixeldetector}.

The dataset assumes a right-handed Cartesian coordinate system is defined with the $z$ axis oriented along the beam axis, the $x$ axis toward the center of the collider, and the $y$ axis oriented upward. 
The $x$ and $y$ axes define the transverse plane, while the $z$ axis identifies the longitudinal direction. 
The radial coordinate is defined as $r=\sqrt{x^2+y^2}$.
The azimuth angle $\phi$ is computed with respect to the $x$ axis in radians in the [$-\pi, \pi$] range. 
The polar angle $\theta$ is measured from the positive $z$ axis and is used to compute the pseudorapidity $\eta = -\log(\tan(\theta/2))$ and the pseudoangular separation $\Delta R = \sqrt{\Delta\phi^2+\Delta\eta^2}$.
The transverse momentum ($\pt$) is the projection of momentum on the ($x$, $y$) plane. 
We use natural units such that $c=1$ and express energy and momentum in units of electronvolt (eV).

\subsection{Graph construction} 

\label{sec:Graph}
Each event's tracker hits are encoded into a \textit{hitgraph} based on a set of selection criteria applied to the candidate nodes and edges.
A set of truth filters are applied to hits before they are assigned to graph nodes. 
In particular, a \textit{$\pt$ filter} rejects hits generated by truth particles with $\pt$ below some minimum threshold \ptmin, a \textit{noise filter} rejects hits generated by noise (no associated truth particle), and a \textit{same-layer filter} rejects all but one hit per layer for each truth particle. 
After this initial hit filtering yields a reduced set of nodes, we choose to connect certain nodes with edges which are most likely to represent true track segments based on geometric considerations.
These chosen edges are a superset of all possible edges that can be classified, therefore it's important to accept as many true track segments as possible, i.e. ensure high \textit{efficiency}, defined as the ratio of true track segment edges contained in the graph to the total number of possible true track segments. 
Conversely, permitting too many edges at this stage, such as a fully-connected hitgraph with $\frac{1}{2}\nnodes(\nnodes-1)$ edges, can result in a highly inefficient or intractable computation problem, due to low \textit{purity}, defined as the number of true track segment edges divided by the total number of edges in the graph. 
This represents a fundamental trade-off between different edge construction algorithms: they must simultaneously maximize efficiency without minimizing purity.

In this work, we use a purely geometric graph construction algorithm that determines whether or not to create an edge with features $a_{ij}$ between hits $i$ and $j$. 
Only pixel detector hits are considered, pseudorapidity is restricted to $\eta\in[-4,4]$, and the noise and same-layer hit filters are applied. 
The same-layer filter introduces an ambiguity in defining edges between the barrel and innermost endcap layers. 
Specifically, barrel hits generated by the same particle could produce multiple true edges incoming to a single endcap hit. 
The resulting triangular edge pattern conflicts with the main assumption of the same-layer filter, that only one true track segment exists between each subsequent layer. 
Thus, a \textit{barrel intersection} cut is applied, in which edges between a barrel layer and an innermost endcap layer are rejected if they intersect with any intermediate barrel layers. 
In addition to the barrel intersection cut, edges must also satisfy constraints on the geometric quantities $z_0=z_i-r_i\frac{z_j-z_i}{r_j-r_i}$ and $\Delta\phi / \Delta r=\frac{\phi_j-\phi_i}{r_j-r_i}$. 
The restrictions are $z_0 < 15$\unit{cm} and $\Delta\phi / \Delta r < 0.0006$ and $\pt^{\mathrm{min}} = 2\GeV$. 

\begin{figure*}[t]
    \begin{center}
    \includegraphics[width=0.49\textwidth]{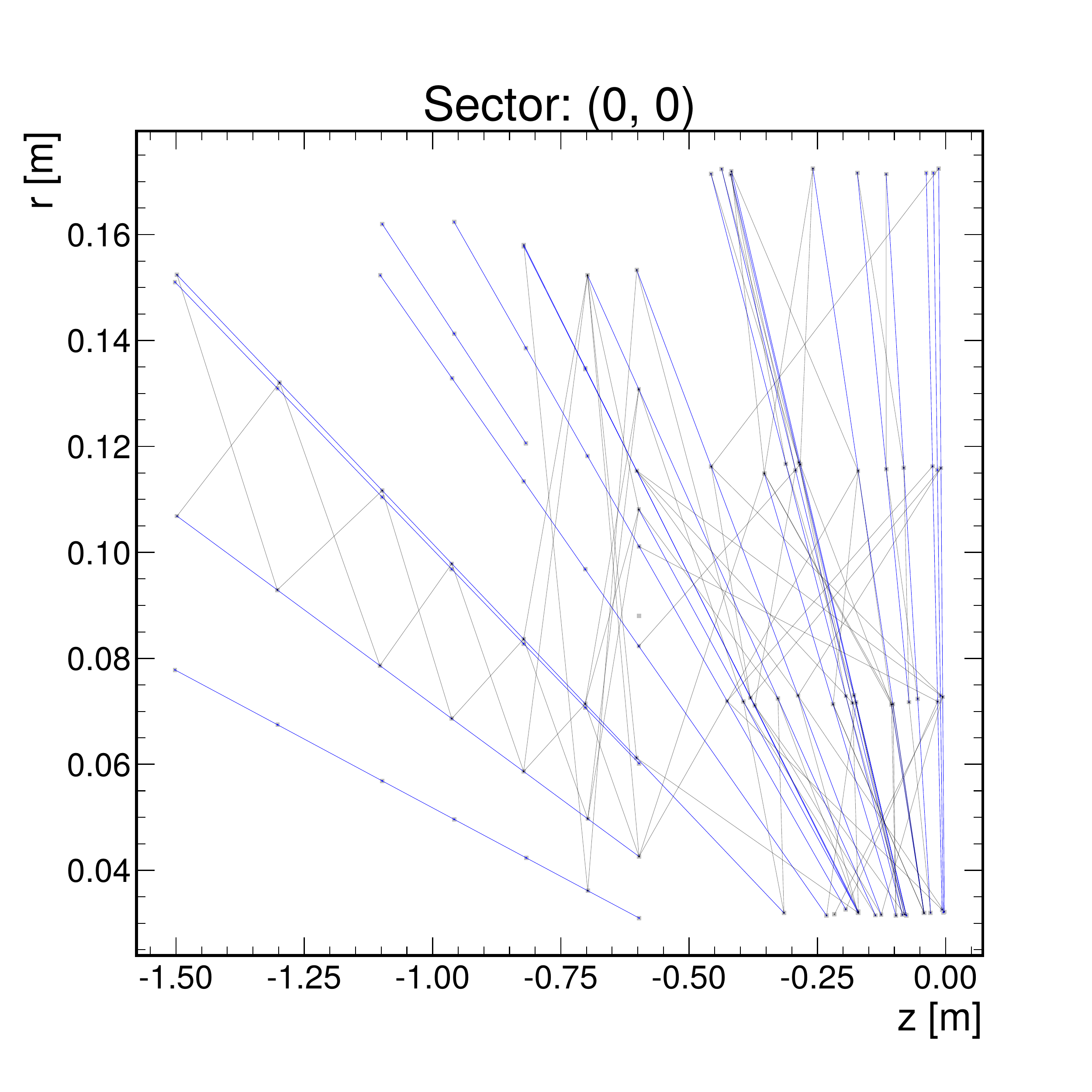}
    \includegraphics[width=0.49\textwidth]{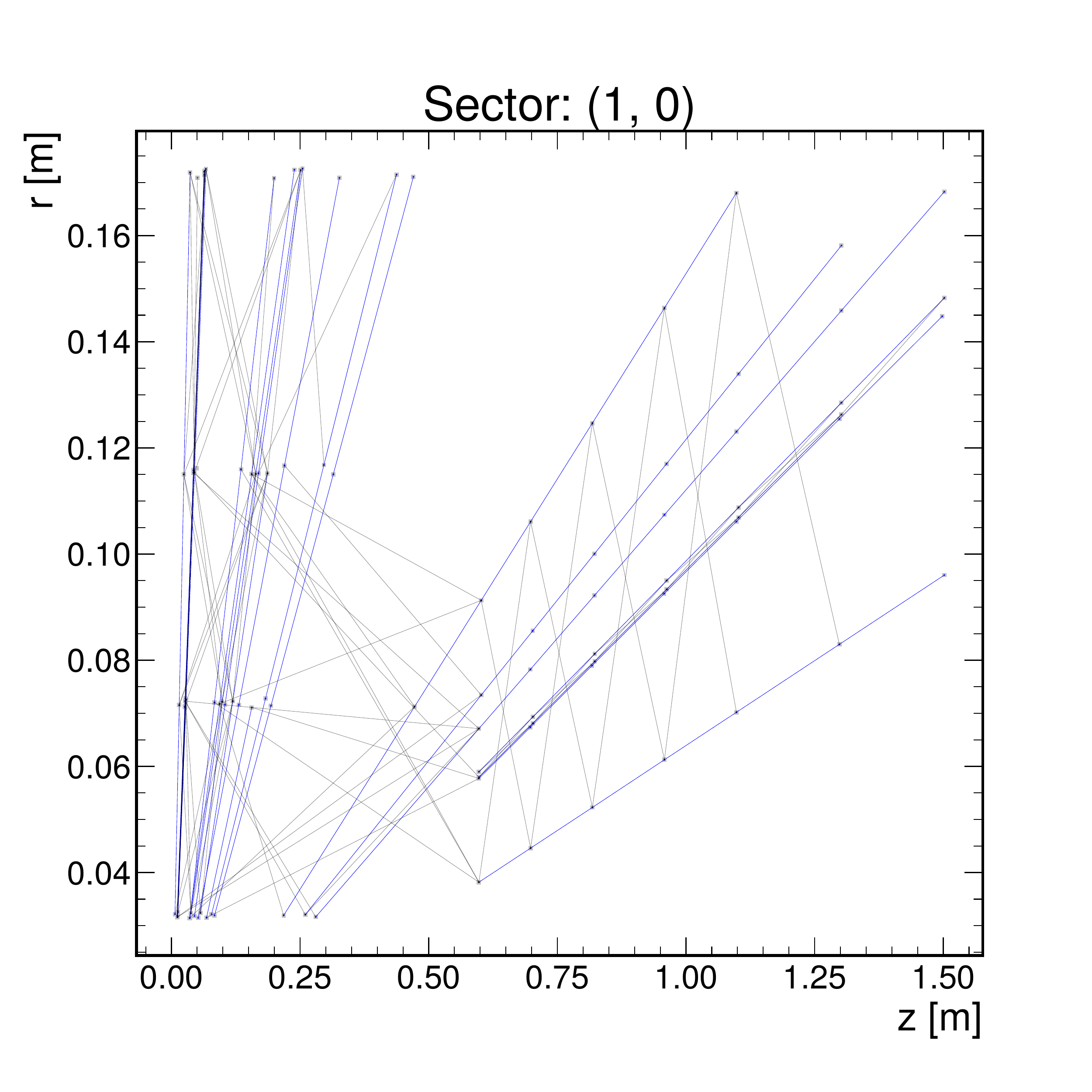}
    \end{center}
    \caption{Example graphs showing 2 of the sectors for one event with $\ptmin = 2\GeV$, $z_0 < 15$\unit{cm}, $\Delta\phi / \Delta r < 0.0006$, 8 $\phi$ sectors, and 2 $\eta$ sectors.
    True track segments are denoted by blue edges, while false track segments are denoted by gray.}
    \label{fig:example_graphs}
\end{figure*}

Even after these geometric and truth particle $\pt$ restrictions, a single TrackML event can feature thousands of nodes and edges.
In order to keep the graph sizes manageable for resource-constrained environments, we can subdivide each event graph into a certain number of $\phi$ and $\eta$ sectors depending on the latency and resource constraints of the system and the range of applicability of the given implementation.
Example graphs for 2 different sectors in one event when subdividing an event into 8 $\phi$ sectors and 2 $\eta$ sectors are shown in Fig.~\ref{fig:example_graphs}.

\begin{figure*}[t]
    \begin{center}
    \includegraphics[width=0.49\textwidth]{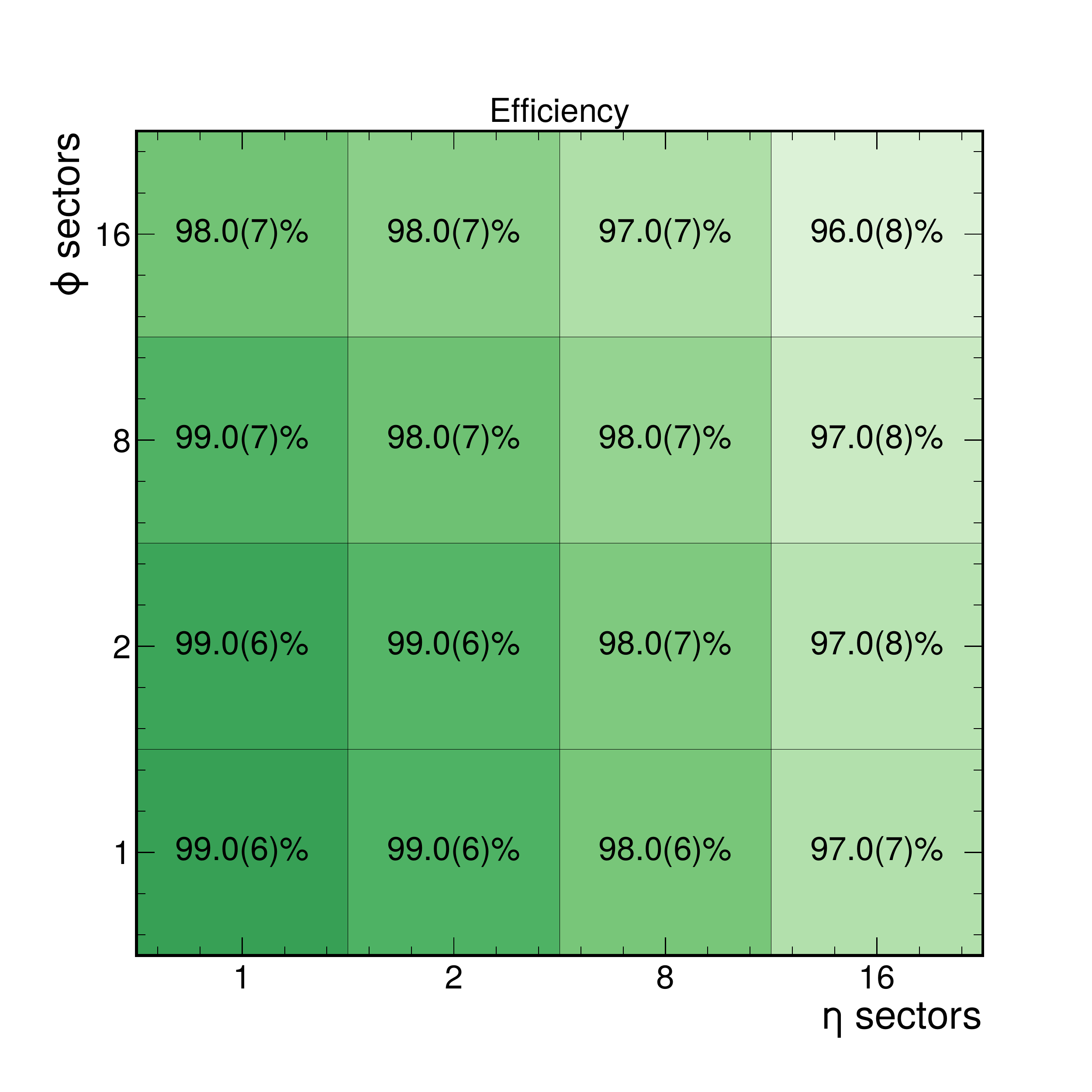}
    \includegraphics[width=0.49\textwidth]{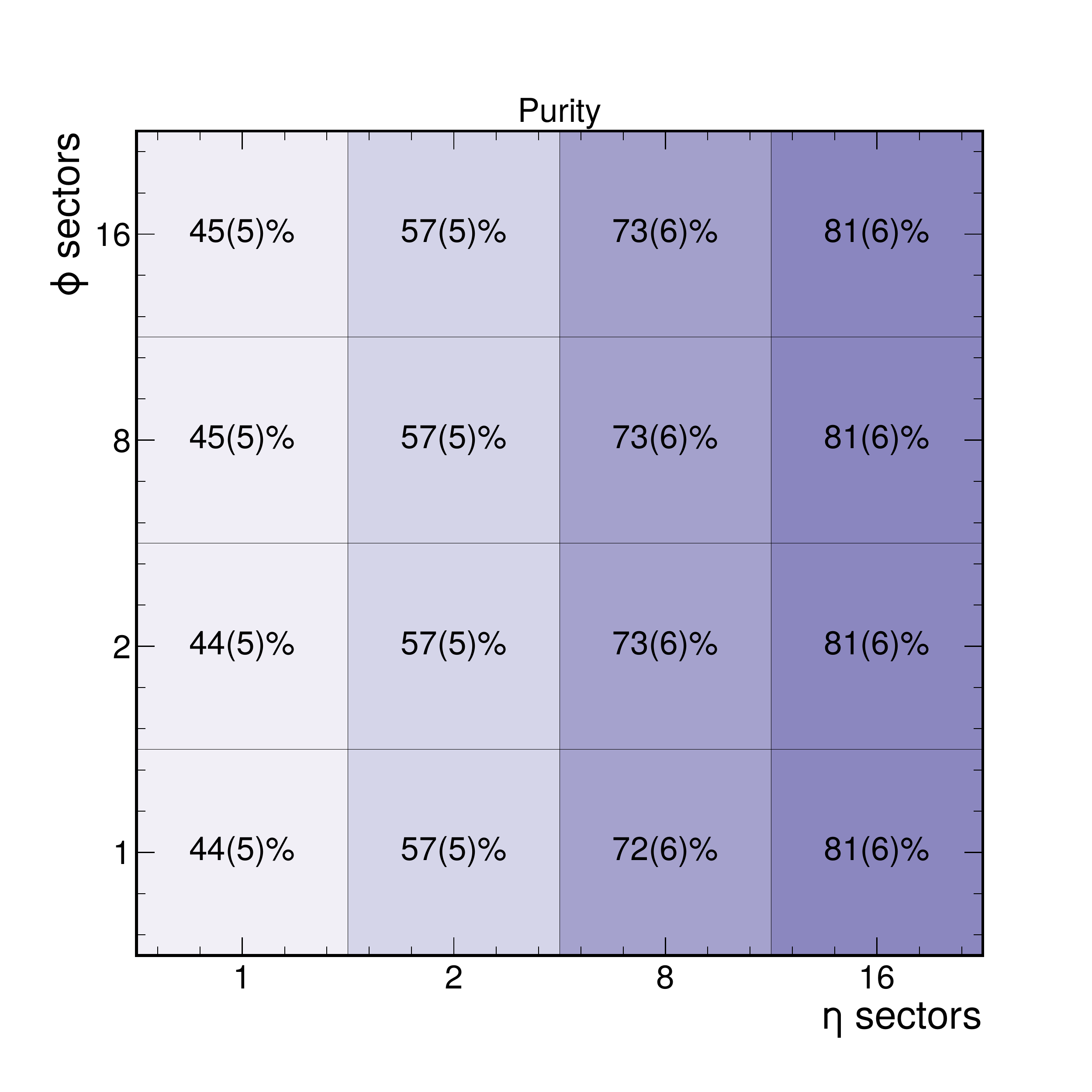}
    \end{center}
    \caption{
    Efficiency (left) and purity (purity) of the hitgraphs studied for different numbers of $\eta$ and $\phi$ sectors based on 50 events in \texttt{train\_1}.}
    \label{fig:efficiency_purity}
\end{figure*}

Based on these graph construction criteria, we can measure the efficiency and purity of the resulting graphs.
These are shown for different choices for the number of $\phi$ and $\eta$ sectors in Fig.~\ref{fig:efficiency_purity} based on 50 events in \texttt{train\_1} TrackML sample.
In particular for 8 $\phi$ sectors and 2 $\eta$ sectors, the graphs retain an efficiency of 98\% and a purity of 57\%.

In addition to the need to subdivide the graphs to reduce size, for a fixed-latency FPGA implementation, hardware resources cannot be dynamically reallocated to accept variable-size input arrays, so the graph sizes must be made uniform.
Thus, we typically consider a fixed maximum input graph size.
One common choice is to truncate the graphs based on the 95th percentile graph size for each sector\footnote{We note that for up to 2 $\eta$ sectors, each of the sectors has the same hit and track multiplicity distribution because the multiplicity depends on $|\eta|$ but not on $\phi$. 
In general, for a larger number of $\eta$ sectors, a variable maximum graph size could be chosen depending on the $|\eta|$ range of the sector.}.
Thus, only 5\% of the input graphs will be truncated.
For smaller graphs that have fewer nodes or edges, we zero-pad the feature matrices to create `null' nodes and add connections between `null' nodes to create `null' edges.
We discuss the effects of the graph truncation and zero-padding on the network performance in Sec.~\ref{sec:results}.

\begin{figure*}[!htbp]
\begin{center}
  \includegraphics[width=0.49\textwidth]{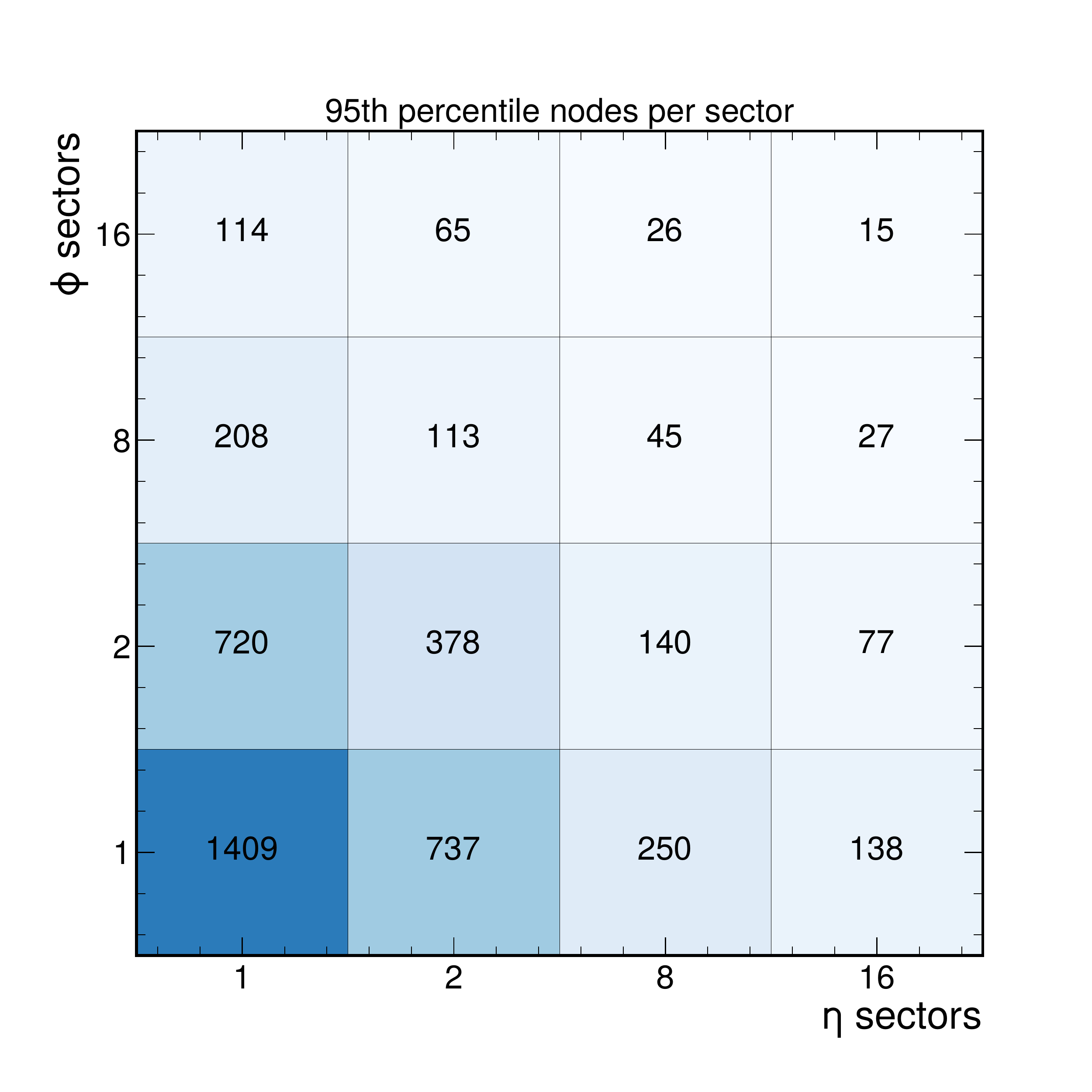}
  \includegraphics[width=0.49\textwidth]{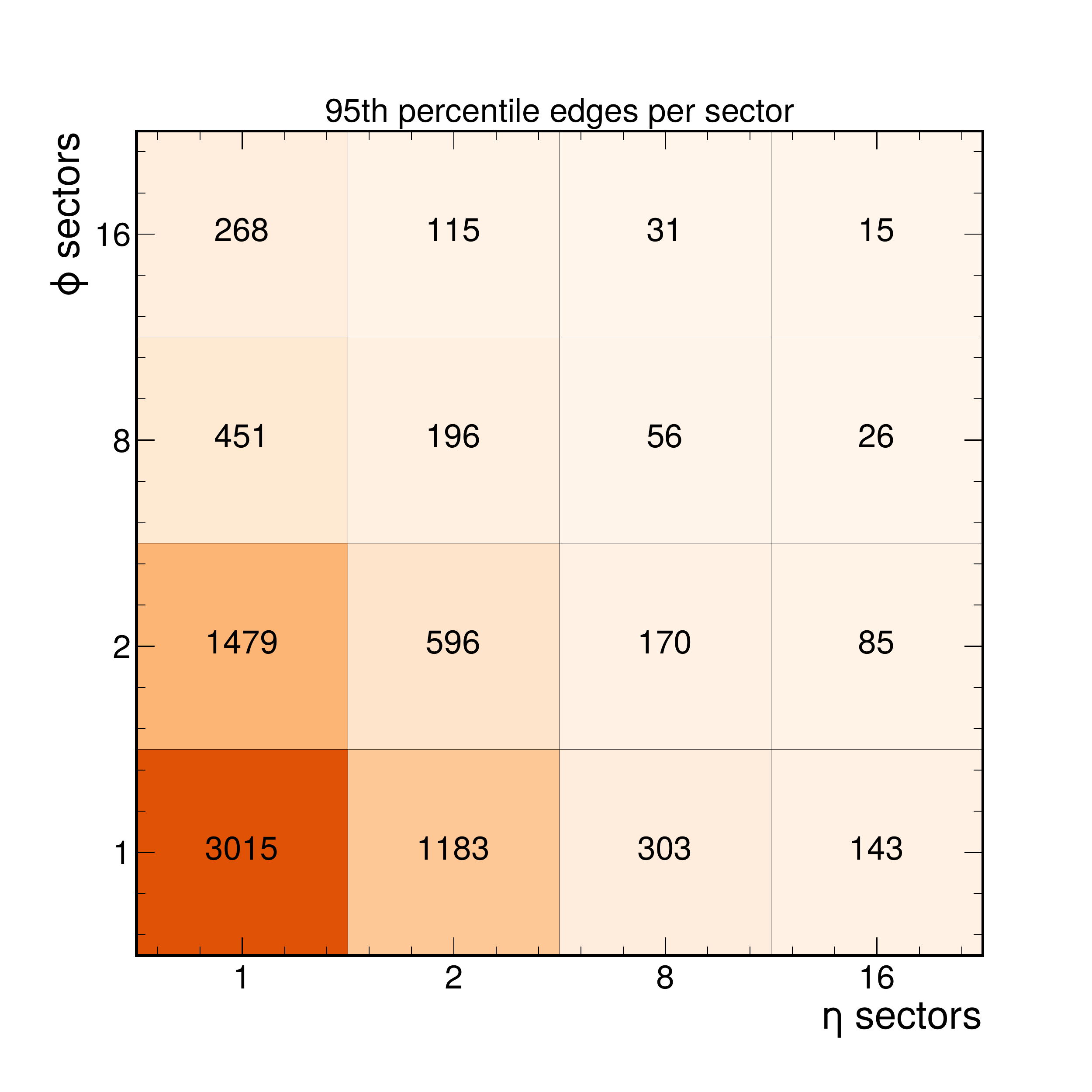}
  \end{center}
\caption{95th percentile of the number of nodes and edges in each sector for the 2\GeV graphs as a function of the number of $\eta$ and $\phi$ sectors, based on 50 events in \texttt{train\_1}.}
  \label{fig:graphs}
\end{figure*}

Figure~\ref{fig:graphs} shows the 95th percentile for the number of nodes and edges in each sector depending on the number of sectors chosen.
For example, the 95th percentile graph size for 8 $\phi$ sectors and 2 $\eta$ sectors is 113 nodes and 196 edges for this 2\GeV graph construction.
Depending on the range of applicability for a given FPGA implementation, a different graph construction and segmentation strategy can be adopted.
In particular, if a more relaxed set of graph construction criteria is adopted, a greater number of nodes and edges will be included per event.
In the Supplementary Material, we demonstrate how the number of nodes and edges vary when considering 1\GeV graphs. 
In particular, the 95th percentile for the number of nodes and edges per event is nearly 6,500 and 20,000, respectively.
To accommodate these denser hitgraphs in resource-constrained implementations, a finer segmentation is required.

\section{Interaction network}
\label{sec:IN}

Fundamentally, GNNs produce new graph embeddings, leveraging them to produce graph-level, edge-level, or node-level predictions.
One iteration of an IN contains two ``update'' functions, $\phi$, and an ``aggregation'' operation, which, for simplicity, we take to be a simple summation.

The \textit{edge block} computes a four-dimensional output $a^{\prime}_{ij}$ for each edge, known as the updated edge feature or ``message.''
\begin{align}
    a^{\prime}_{ij} &= \phi_{R,1}(x_i, x_j, a_{ij})~, 
  \label{eq:edgeblock}
\end{align}
where $x_i$ and $x_j$ are the input features of node $i$ and $j$, respectively, which are the ($r$, $\phi$, $z$) coordinates of the corresponding hit, and $a_{ij}$ is the set of input edge features ($\Delta r$, $\Delta \phi$, $\Delta z$, $\Delta R$).
These messages are subsequently aggregated (summed) over the corresponding connected nodes belonging to the neighborhood $\mathcal{N}(i)$ of node $i$.
\begin{align}
\overline{a}^{\prime}_{i} &= \sum_{j\in \mathcal{N}(i)}a_{ij}^{\prime}
\end{align}
These two steps are known as the message-passing operation. 
The aggregation function, here taken to be summation, maps edge-specific information to node-specific outputs by compiling information based on the connected node indices.
To apply generically to unordered graph-structured data, the chosen aggregation function must be invariant to permutations of their inputs, and should take variable numbers of arguments.
Other examples include an element-wise mean, maximum, and minimum.
This construction ensures permutation equivariance of the GNN for edge- and node-level outputs.

The \textit{node block} computes a three-dimensional output for each node
\begin{align}
 x_i^{\prime} &= \phi_O(x_i, \overline{a}_{i}^{\prime})
\end{align}
that can be thought of as an update of the node features, which takes into account the previous node features, and one round of message passing among neighboring nodes.

An additional partial edge block gives the final one-dimensional \textit{edge weights}
\begin{align}
    a_{ij}^{\prime\prime} = \phi_{R,2}\big(x_i^{\prime}, x_j^{\prime}, a_{ij}^{\prime}\big),\label{eq:edgeweight}
\end{align}
In this way, we produce edge weights from the re-embedded graph with node features and edge features.

\begin{figure*}[t]
\begin{center}
  \includegraphics[width=\textwidth]{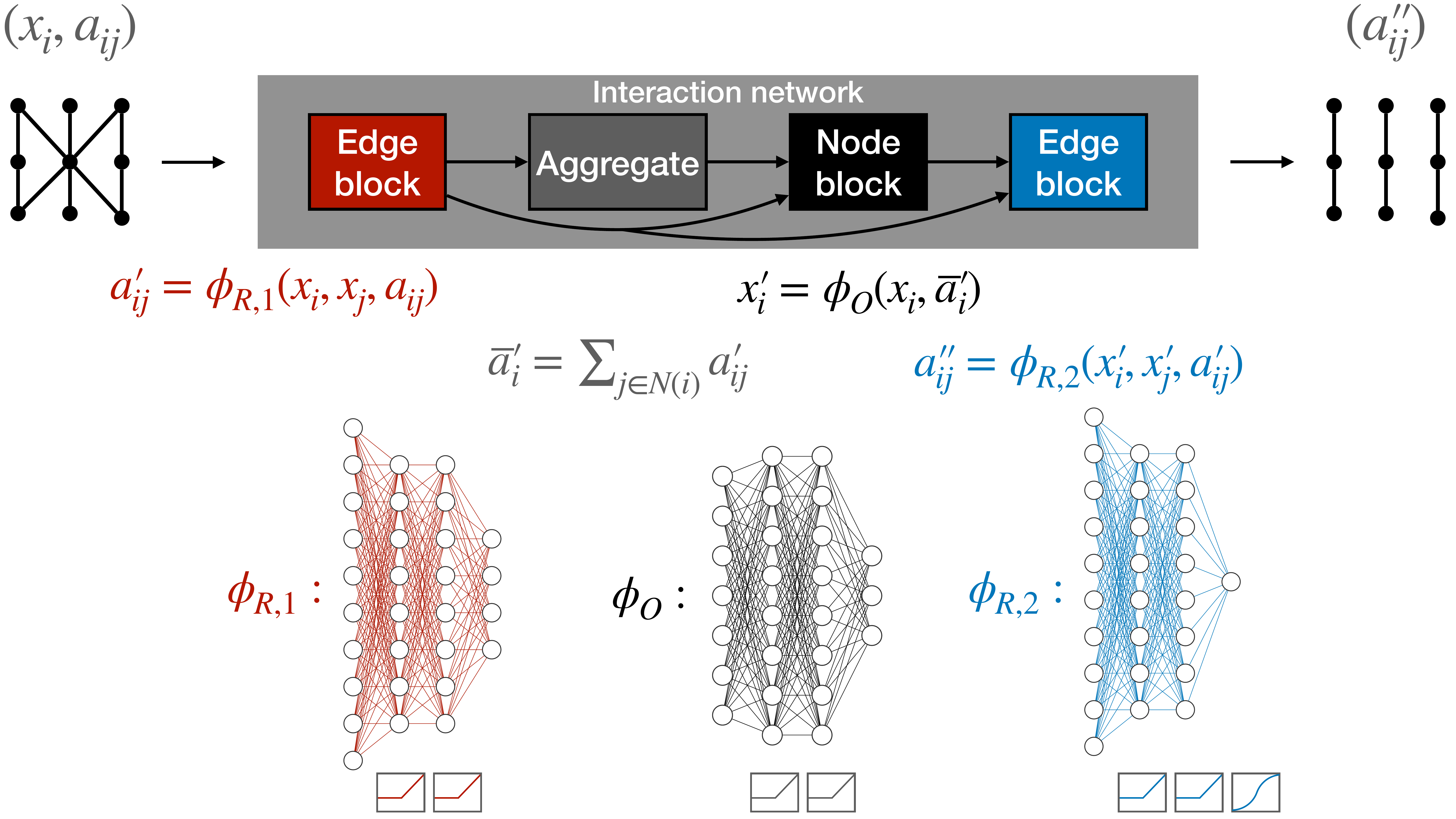}
  \end{center}
\caption{The complete IN forward pass with the edge, edge-aggregation, and node blocks.
The functions $\phi_{R,1}$, $\phi_{R,2}$, and $\phi_O$ are multilayer perceptrons (MLPs) with rectified linear unit (ReLU) activation functions on the hidden layers. 
In addition, the $\phi_{R,2}$ outputs have a sigmoid activation.}
  \label{fig:network_diagram}
\end{figure*}

The full forward pass, comprised of edge and node blocks used to predict edge weights, is shown in Fig.~\ref{fig:network_diagram}.
The functions $\phi_{R,1}$, $\phi_{R,2}$, and $\phi_O$ are multilayer perceptrons (MLPs) with rectified linear unit (ReLU) activation functions~\citep{relu1,relu2} on the hidden layers. 
The ReLU activation function behaves as an identity function for positive inputs and saturates at 0 for negative inputs. 
Notably, the $\phi_{R,2}$ outputs have sigmoid activations and we optimize the binary cross-entropy (BCE) loss between the truth targets and the outputs, such that the resulting edge weights $a_{ij}^{\prime\prime}\in[0,1]$ can be interpreted as independent probabilities that each edge is a track segment.
However, we note that this probabilistic interpretation is not required to build a tracking algorithm based on this IN.

Throughout the following studies, the architecture in Fig.~\ref{fig:network_diagram} is held at a constant size of 528 trainable parameters, corresponding to 8 hidden units (h.u.) per layer in each of the MLPs.
Assuming every MLP layer has the same number of h.u., 8 h.u. per layer is sufficient to recover the maximum classification accuracy with models trained on $\ptmin=2\GeV$ segmented graphs.
In the following studies, models are trained on graphs built with $\ptmin=2\GeV$. 
A total of about 28\,k graph sectors (corresponding to 1770 total events and 16 sectors per event) belonging to the TrackML \texttt{train\_1} sample are randomly divided into 70\% for training, 20\% for validation, and 10\% for testing.
The Adam optimizer is used to facilitate training~\citep{adam}. 
It is configured with a learning rate of $10^{-3}$ and $\epsilon=10^{-8}$, and the model is trained for 1 epoch, which we find is enough time for the model to reach convergence.

\section{\hlsfml User Interface}
\label{sec:hls4ml}

The \hlsfml workflow performs automatically the task of translating a trained NN, specified by the model's architecture, weights, and biases, into the specification of a hardware accelerator.
Because every application is different, the goal of the \hlsfml package is to empower the user to perform this optimization through automated NN translation and design iteration.
\hlsfml leverages HLS to generate hardware modules from code written in high-level programming languages like \textsc{C}/\textsc{C++}~\citep{numan2020towards}.
Although it may lead to slightly less optimal performance than RTL-based design, HLS-based design has significant benefits: it raises the level of abstraction, reduces the iteration time, simplifies the validation phase, and enables greater exploration and evaluation of design alternatives.

\begin{figure*}[t]
\begin{center}
\includegraphics[width=0.99\textwidth]{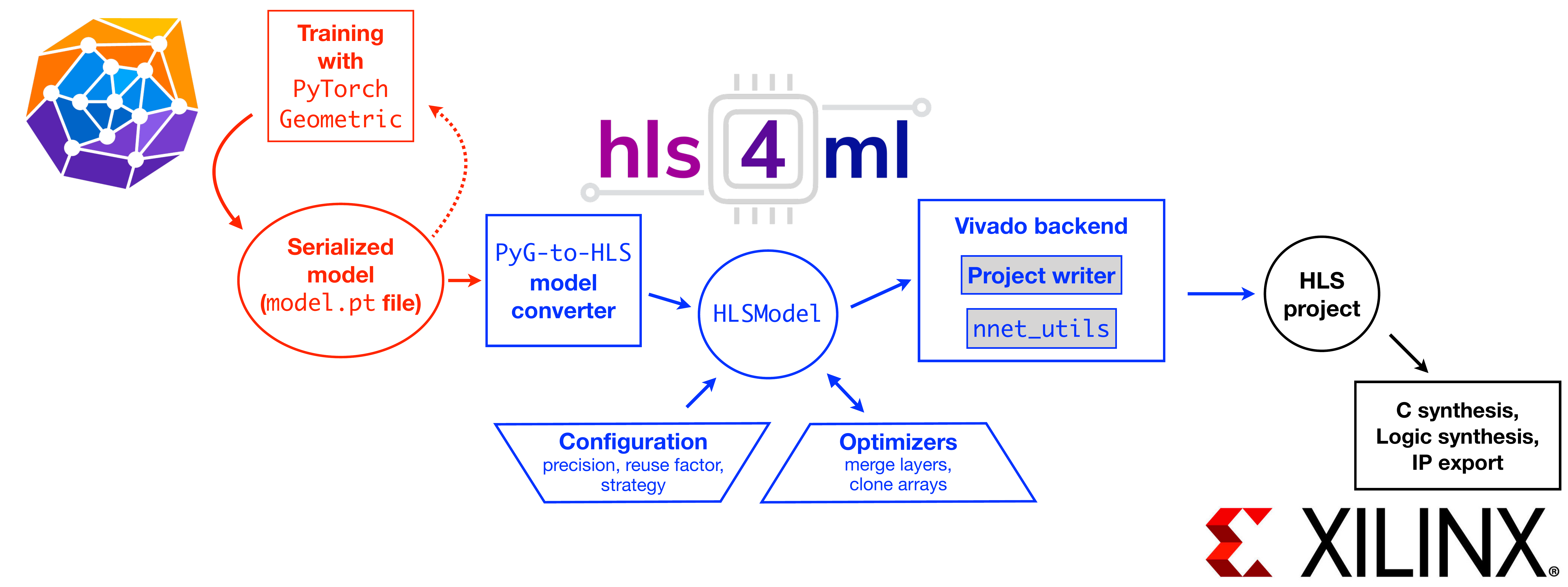}
\end{center}
\caption{The workflow to translate a GNN model into an FPGA implementation using \hlsfml.
The red boxes (left) describe the model training steps performed within ML software frameworks like \textsc{PyTorch Geometric}.
The \hlsfml configuration and conversion steps are shown in the blue boxes (center).
A model converter translates the model from \textsc{PyTorch Geometric} into an intermediate \textsc{HLSModel} representation.
This representation can be further configured and optimized.
Different backend writers can be used to export the model into a given vendor-specific language, such as Vivado HLS.
The black boxes (right) illustrate possible ways to export and integrate the HLS project into a larger hardware design for example for Xilinx FPGAs.}
\label{fig:flow}
\end{figure*}

Figure~\ref{fig:flow} shows the schematic of a typical workflow.
The first part of the workflow illustrated in red depicts the model training and optimization performed with \textsc{PyTorch Geometric} (\textsc{PyG})~\citep{PyTorchGeometric}.
The blue section of the workflow is performed with \hlsfml, which translates the model into an HLS project that can subsequently be synthesized and implemented on an FPGA or ASIC, as depicted by the black section.

The \hlsfml workflow is built on the concept of independent NN ``layers'' that process some input data to produce some output data.
Each layer and activation type is implemented as a separate configurable module customized to perform that specific operation.
During the \hlsfml conversion, these modules are composed in the correct way to perform the inference for a full ML model.

Specifically, this conversion step automatically generates a top-level C++ executable that uses predefined C++ templates representing different NN layers provided through the \textsc{nnet\_tools} library in \hlsfml.
For this GNN application, the edge block, node block, and edge-aggregation block are each considered separate layers.
\hlsfml also provides a number of configurable parameters which can help the user explore and customize the space of latency, throughput, power, and resource usage tradeoffs for their application.

To provide flexibility and ease-of-use, \hlsfml provides both a programming API and visualization capabilities.
Figure~\ref{fig:flow} also illustrates the internal structure of the \hlsfml \textsc{Python} package.
The package first converts the user-specified model into a common internal representation of the network graph.
Converters exist for \textsc{Keras}, \textsc{TensorFlow}, \textsc{PyTorch}, and \textsc{ONNX} model formats.
For this work, a new converter for the GNN model provided as a \textsc{PyTorch Geometric} model was added. 
At the conversion step, the user-provided configuration is also attached to the model. 
We note additional user-provided information is required: the order of the modules.


One key feature of the programming API is the capability to execute the bit-accurate emulation of the generated HLS-synthesizable code in the \textsc{Python} environment, for example as a Jupyter Notebook.
The \hlsfml API enables a workflow where inference can be performed on \textsc{NumPy}~\citep{harris2020array} array data in \textsc{Python}.
In addition to evaluating the \hlsfml model output, users can access the detailed output of any layer of the network, which can aid in debugging and performing hyperparameter optimization for quantized models.
For this work, the capability of running inference on models with multiple inputs (node attributes, edge attributes, and the edge index) was added.

\section{FPGA HLS Implementations}
\label{sec:FPGA}

To meet the strict latency and throughput requirements of the L1T, we developed a \textit{throughput-optimized} design, in which the processing of multiple input graphs overlaps in time.

Conversely, we also consider a \textit{resource-optimized} design, which requires more clock cycles but also consumes far fewer resources.
This design has the advantage that it scales to larger input graphs.

Tests of the \hlsfml implementation target a Xilinx Virtex UltraScale+ VU9P FPGA (part number \texttt{xcvu9p-flga2104-2L-e}) with a 5\unit{ns} clock period.
The target FPGA has 6,840 digital signal processing slices (DSPs), 1,182,240 look-up tables (LUTs), 2,364,480 flip-flops (FFs), and 75.9\unit{Mb} of block random access memory (BRAM)~\citep{datasheet}.
As discussed in Sec.~\ref{sec:Graph}, the input graph size can be adjusted by segmenting the detector more finely in $\eta$ and $\phi$ sectors.
We find that large input graph sizes prohibit the use of the throughput-optimized design. 
Therefore, to make scans of resources and timing tractable, we further synthesize designs for graphs of up to 28 nodes and 56 edges.
For the resource-optimized design, we also demonstrate how the design scales to larger graphs by presenting results for up to 1,344 nodes and 2,688 edges.
For all results shown, we fix the ratio of the number of edges to nodes to 2, as we empirically find that this is close to the observed ratio for the 2\GeV graphs we consider.

\subsection{HLS preprocessor directives}
The FPGA HLS implementation is written using Vivado HLS~\citep{vivado20192} and integrated with the transpiler \hlsfml. 
Vivado HLS makes use of preprocessor directives, called \textit{pragmas}, in order to specify certain high-level design choices. 
A full description of the two different designs requires an understanding of some of these pragmas, namely \texttt{ARRAY\_PARTITION}, \texttt{UNROLL}, and \texttt{PIPELINE}.

Applying \texttt{ARRAY\_PARTITION} to a given array will partition it into several smaller arrays by one of three different methods. 
Block partitioning with a factor of $N$, will partition the array into $N$ consecutive blocks. 
Cyclic partitioning with a factor of $N$, will create $N$ smaller arrays by interleaving elements from the original array such that the $i$th smaller array contains elements $i, i+N, i+2N, \dots$ from the original array. 
Complete partitioning will fully decompose an array into its individual elements. 
The benefit of partitioning is that different functions can concurrently access and operate on the separate, smaller arrays, but this comes at the cost of utilizing more memory registers.

The \texttt{UNROLL} pragma is used within for-loops. 
For a for-loop that normally has $M$ iterations, fully unrolling this loop will create $M$ physical copies of the for-loop logic, so that each iteration can run concurrently.
Partially unrolling this loop with a factor of $N<M$ will create $N$ copies of the for-loop, each of which will iterate $\lceil M/N\rceil$ times; Vivado HLS does not require that $M$ is an integer multiple of $N$. 
Like partitioning, unrolling sacrifices some resources in exchange for lower latency.

Applying the \texttt{PIPELINE} pragma on a function will minimize the function's initiation interval (II), which is the wait time required before a new input can be processed. 
Pipelining a function allows it to accept a new input before it has finished operating on an earlier input. 
For example, consider a simple three-layer MLP, and two different inputs for the MLP: A and B.
Typically, one would send input A into the MLP, wait for it to pass through all three layers, and then send input B into the MLP.
With pipelining, one can send input A through the first layer of the MLP, and then send input B through the first layer as soon as input A reaches the second layer.
Likewise, the second layer can accept input B once input A has made it to the third layer.
With pipelining, idle resources are used in order to minimize II without requiring any additional resources. 
The amount of pipelining is limited by the logic and timing of the relevant design.
For example, it is possible to pipeline a for-loop as long as each iteration is independent.
\vspace{1ex}
\subsection{Throughput-optimized design}
\label{sec:throughput_design}
In the throughput-optimized design, pipelining is performed at the level of the IN edge block, node block, and edge-aggregation block, and the amount of pipelining is tunable through the \textit{reuse factor} (RF) parameter, which controls the II of each block.
In conjunction with block-level pipelining, all loops are fully unrolled to decrease latency and all arrays are completely partitioned to allow concurrent access to each element.
These are implemented through the HLS pragmas described above.
Fully unrolling the arrays places a constraint on how large the input graphs can be, given the limitations imposed by the Vivado HLS compiler.
In practice, we find graphs of 28 nodes and 56 edges are near the maximum we can consider with this design.

The edge block receives as input three fully partitioned arrays: the node feature matrix with arbitrary fixed-point precision, the edge feature matrix with arbitrary fixed-point precision, and the edge index matrix with an arbitrary integer precision. 
The output of the edge block is the updated matrix of edge features.
The first step is to retrieve the appropriate pair of node features for each edge through the edge index. 
This is done through a non-static array index.
This pair of node features is concatenated with the corresponding edge features in a temporary array.
Subsequently, a fully-connected NN (of up to four layers in depth) is applied to compute the updated edge features. 
HLS pseudocode illustrating the main structure and pragmas of the edge block is shown in Algorithm~\ref{alg:edgeblock_throughput}.

\begin{algorithm}
\caption{Throughput-optimized edge block.}
\label{alg:edgeblock_throughput}
\begin{algorithmic}
\Require node\_attr[$\nnodes$][node\_dim], edge\_attr[$\nedges$][edge\_dim], edge\_index[$\nedges$][2]
\Ensure edge\_update[$\nedges$][edge\_dim]
\State pipeline algorithm with factor RF \Comment{HLS pragma}
\Procedure{partition arrays}{node\_attr, edge\_attr, edge\_update}
\State completely partition node\_attr \Comment{HLS pragma}
\State completely partition edge\_attr \Comment{HLS pragma}
\State completely partition edge\_update \Comment{HLS pragma}
\EndProcedure
\Procedure{Create NN input}{node\_attr, edge\_attr, edge\_index}
\State initialize phi\_input[$\nedges$][edge\_dim$+2\times$node\_dim]
\State completely partition phi\_input \Comment{HLS pragma}
\For{$i\gets 1, \nedges$}
\State completely unroll loop  \Comment{HLS pragma}
\State receiver\_index $\gets$ edge\_index[$i$][0]
\State sender\_index $\gets$ edge\_index[$i$][1]
\State receiver $\gets$ node\_attr[receiver\_index]
\State sender $\gets$ node\_attr[sender\_index]
\State edge $\gets$ edge\_attr[$i$]
\State phi\_input[$i$] $\gets$ $\langle$receiver, sender, edge$\rangle$
\EndFor
\State \textbf{return} phi\_input
\EndProcedure
\Procedure{Compute edge update}{phi\_input}
\For{$i\gets 1, \nedges$}
\State completely unroll loop  \Comment{HLS pragma}
\State edge\_update[$i$] $\gets$ NN(phi\_input[$i$])
\EndFor
\State \textbf{return} edge\_update
\EndProcedure
\end{algorithmic}
\end{algorithm}

The edge-aggregation block takes as input the updated edge feature matrix and the edge index matrix, and returns the aggregated updated edge features gathered along the receiver nodes.
To ensure a greater design flexibility, three aggregation methods are supported: sum, average, and maximum.
Summation is the simplest aggregation algorithm, and simply requires appropriately summing the corresponding edge features for each receiver node.
Average requires an additional step of counting the number of edges connected to each receiver node, and subsequently dividing by this number. 
The division is implemented through a predefined look-up table.
In maximum aggregation, we initially set the output matrix to the most negative value allowed by the datatype.
Subsequently, we use a tournament method to find the maximum value.
We note that maximum aggregation requires the longest latency and the most resources.
Therefore, summation and average are choices better suited for this FPGA implementation. 

Finally, the node block receives the current node features and the aggregated updated edge features, concatenates them, processes the concatenated product with a NN and returns the updated node features. 
It does not require any non-static array access. 

\vspace{1ex}
\subsection{Resource-optimized design}
\label{sec:resource_design}
In the resource-optimized design, pipelining is performed at both the task-level of the whole design and the level of each functional block, using the RF to determine each loop's II and a \textit{parallelization factor} (PF) to determine the degree of loop-unrolling.
The task-level pipelining allows every functional block to execute concurrently, reducing II of the total design. 
Within each functional block, the PF is used to adjust the number of unrolled loops. 
The resource-optimized design can handle large graph sizes by adjusting PF to fit FPGA resource capacity (higher PF means more unrolled loops and greater resource usage). 
However, the latency will increase in proportion to the graph size and violates the 4\unit{$\mu$s} latency constraint when the graph size is larger than about 896 nodes and 1,792 edges.

The resource-optimized edge block has the same inputs and outputs as the throughput-optimized edge block.
Algorithm~\ref{alg:edgeblock_resource} shows the pseudocode of this design, which adopts different explicit memory access methods to handle arrays that require static or non-static access. 
Static-access arrays (edge features, edge index, and updated edge features) are cyclically partitioned with a factor equal to the PF multiplied by the array's feature dimension (e.g. the edge index array's feature dimension is 2, the edge feature array's feature dimension is the length of the vector used to represent each edge). 
This way, all the parallel computing elements can access these arrays concurrently. 
The same practice with the node features array, which requires non-static access, would cause latency degradation due to stalling cycles when accessing data of different addresses.
Therefore, the node features array is first copied into PF-many duplicates, and each parallel computing element then accesses node data from its dedicated duplicate. 

\begin{algorithm}
\caption{Resource-optimized edge block.}
\label{alg:edgeblock_resource}
\begin{algorithmic}
\Require node\_attr[$\nnodes$][node\_dim], edge\_attr[$\nedges$][edge\_dim], edge\_index[$\nedges$][2]
\Ensure edge\_update[$\nedges$][edge\_dim]
\Procedure{partition arrays}{node\_attr, edge\_attr, edge\_index, edge\_update}
\State cyclically partition node\_attr with factor node\_dim$\times$PF \Comment{HLS pragma}
\State cyclically partition edge\_attr with factor edge\_dim$\times$PF \Comment{HLS pragma}
\State cyclically partition edge\_index with factor 2$\times$PF \Comment{HLS pragma}
\State cyclically partition edge\_update with factor edge\_dim$\times$PF \Comment{HLS pragma}
\EndProcedure
\Procedure{Copy node attributes}{node\_attr}
\State initialize node\_attr\_copies[PF][$\nnodes$][node\_dim]
\For{$i\gets 1, \nnodes$}
\State unroll loop with factor PF \Comment{HLS pragma}
\State pipeline loop with factor RF \Comment{HLS pragma}
\For{$j\gets 1, \mathrm{PF}$}
\State completely unroll loop \Comment{HLS pragma}
\State node\_attr\_copies[$j$][$i$] $\gets$ node\_attr[$i$]
\EndFor
\EndFor
\State \textbf{return} node\_attr\_copies
\EndProcedure
\Procedure{Compute edge update}{edge\_attr, node\_attr\_copies, edge\_index}
\For{$i\gets 1, \nedges$}
\State unroll loop with factor PF \Comment{HLS pragma}
\State pipeline loop with factor RF \Comment{HLS pragma}
\State receiver\_index $\gets$ edge\_index[$i$][0]
\State sender\_index $\gets$ edge\_index[$i$][1]
\State receiver $\gets$ node\_attr\_copies[$i\%$PF][receiver\_index]
\State sender $\gets$ node\_attr\_copies[$i\%$PF][sender\_index]
\State edge $\gets$ edge\_attr[$i$]
\State phi\_input $\gets$ $\langle$receiver, sender, edge$\rangle$
\State edge\_update[$i$] $\gets$ NN(phi\_input)
\EndFor
\State \textbf{return} edge\_update
\EndProcedure
\end{algorithmic}
\end{algorithm}

The resource-optimized edge-aggregation block has the same inputs and outputs as the throughput-optimized edge-aggregation block. 
This block adopts different explicit memory access methods to handle arrays that require static or non-static access. 
Static-access arrays (edge index and updated edge features) are partitioned in the same way as those in the resource-optimized edge block. 
The aggregated edge features array, which requires non-static access, must be handled differently. 
Stalling cycles may arise if there is an attempt to concurrently aggregate two or more edges to the same receiver node. 
Therefore, this block first creates PF-many aggregated edge feature ``duplicates,'' and each parallel computing element then aggregates to its dedicated duplicate. 
Then, these duplicates are themselves aggregated into the final aggregated edge features array.
The corresponding HLS pseudocode is shown in Algorithm~\ref{alg:aggregateblock_resource}.

\begin{algorithm}
\caption{Resource-optimized aggregation block.}
\label{alg:aggregateblock_resource}
\begin{algorithmic}
\Require edge\_update[$\nedges$][edge\_dim], edge\_index[$\nedges$][2]
\Ensure node\_aggr[$\nnodes$][node\_dim]
\Procedure{partition arrays}{edge\_update, edge\_index, node\_aggr}
\State cyclically partition edge\_update with factor edge\_dim$\times$PF \Comment{HLS pragma}
\State cyclically partition edge\_index with factor 2$\times$PF \Comment{HLS pragma}
\State cyclically partition node\_aggr with factor edge\_dim$\times$PF \Comment{HLS pragma}
\EndProcedure
\Procedure{Reset intermediate node features}{node\_interim}
\State initialize node\_interim[PF][$\nnodes$][edge\_dim]
\For{$i\gets 1, \nnodes$}
\State unroll loop with factor PF \Comment{HLS pragma}
\State pipeline loop with factor RF \Comment{HLS pragma}
\For{$j\gets 1, \mathrm{PF}$}
\State completely unroll loop \Comment{HLS pragma}
\State node\_interim[$j$][$i$] $\gets 0$
\EndFor
\EndFor
\State \textbf{return} node\_interim
\EndProcedure 
\Procedure{Aggregate edge update to receiver}{edge\_update, node\_interim, edge\_index}
\For{$i\gets 1, \nedges$}
\State unroll loop with factor PF \Comment{HLS pragma}
\State pipeline loop with factor RF \Comment{HLS pragma}
\State receiver\_index $\gets$ edge\_index[$i$][0]
\State node\_interim[$i\%$PF][receiver\_index] $\gets$ node\_interim[$i\%$PF][receiver\_index]  + edge\_update[$i$]
\EndFor
\State \textbf{return} edge\_update
\EndProcedure
\Procedure{Sum intermediate node features}{node\_interim}
\For{$i\gets 1, \nnodes$}
\State unroll loop with factor PF \Comment{HLS pragma}
\State pipeline loop with factor RF \Comment{HLS pragma}
\For{$j\gets 1, \mathrm{PF}$}
\State completely unroll loop \Comment{HLS pragma}
\State node\_aggr[$i$] $\gets$ node\_interim[$j$][$i$] + node\_aggr[$i$] 
\EndFor
\EndFor
\State \textbf{return} node\_aggr
\EndProcedure
\end{algorithmic}
\end{algorithm}

The resource-optimized node block also has the same inputs and outputs as the throughput-optimized node block. 
It does not require any non-static array access, so all arrays are partitioned in the same manner way that the edge block and edge-aggregation block partition the static-access arrays.   

Finally, in the resource-optimized design, arrays which are used by more than one functional block, or \hlsfml layer, are cloned in order to allow concurrent access. For example, it can be seen in Figure~\ref{fig:network_diagram} that the output of the first edge block is used in the edge-aggregation block, the node block, and the second edge block. 
Therefore, three clones are created from the output of this first edge block, one for each function that uses it. 
This way, no single array is ever used by more than one function. 
To perform this cloning, we developed an \hlsfml cloning layer and a frontend \hlsfml optimizer which searches a design for reused arrays and appropriately inserts instances of this cloning layer wherever necessary.

In summary, the main differences between the resource-optimized and throughput-optimized design are (1) pipelining both at the task-level and within each functional block and (2) duplicating arrays for parallel, concurrent access. 
Although (2) demands more storage space for data, the memory resources of the FPGA used are more than sufficient to support this design choice for all the graph sizes discussed here.



\section{Results}
\label{sec:results}

\begin{figure*}[!htbp]
\begin{center}
  \includegraphics[width=0.49\textwidth]{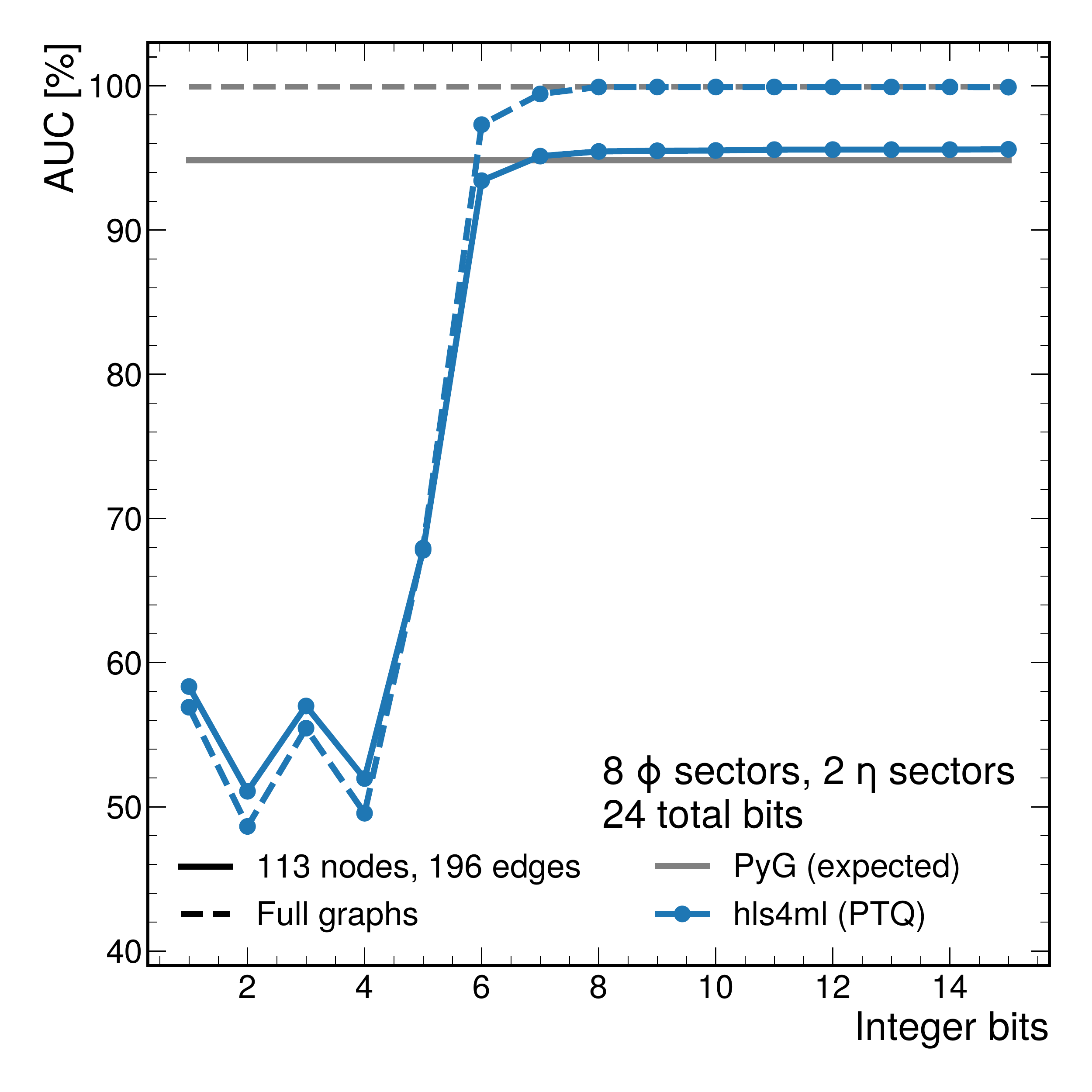}
  \includegraphics[width=0.49\textwidth]{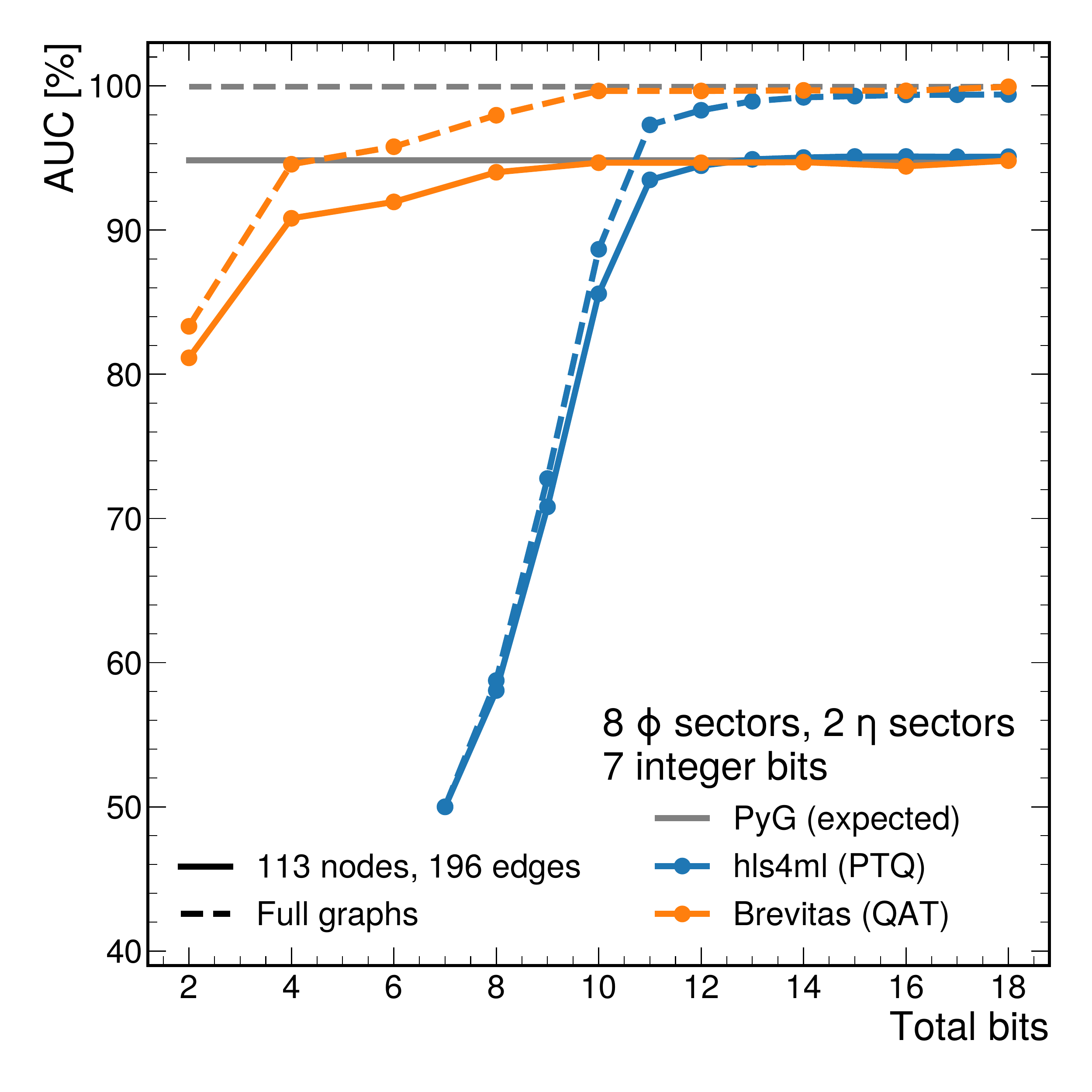}
  \end{center}
\caption{AUC values as a function of the integer bit width with total bit width fixed to 24 (left) and total bit width with integer bit width fixed to 7 (right).
Either the full sectorized 2\GeV graphs (dashed line) or those truncated at 113 nodes and 196 edges (solid line), corresponding to the 95\% percentile graph size, are used as input.
The performance is evaluated with 1000 graphs from \texttt{train\_2}.
With precision greater than \apfixed{12}{7}, the AUC closely approximates the full floating point model for the same graphs.
}
  \label{fig:ROCAUC}
\end{figure*}

For the \hlsfml implementation, we scan the fixed-point precision to determine its impact on the physics performance of the algorithm as well as the latency and resource usage on the FPGA. 
We evaluate the receiver operating characteristic (ROC) curve for the segment classifier, and use the area under the curve (AUC) as a performance metric.
In particular, we first scan the number of integer bits \texttt{Y} when using the Vivado arbitrary precision data type \apfixed{X}{Y}, i.e. \texttt{X} total bits are used and \texttt{Y} bits are used for the integer part (including sign).
From this, we determine that the minimum number of integer bits required to reach the full AUC is 7.
Next, we scan the number of total bits, holding the number of integer bits fixed to 7. 
Figure~\ref{fig:ROCAUC} shows the AUC as a function of the number of integer bits (left) and total bits (right).
We see that with 12 total bits and 7 integer bits, we effectively reproduce the 32-bit floating point model.

With \hlsfml, we employ post-training quantization (PTQ), meaning the model training does not take into account the expected reduced precision.
The required bit width for full performance can be reduced further through techniques like \textit{quantization-aware training} (QAT)~\citep{Coelho:2020zfu,qkeras,Hawks:2021ruw}, in which the effects of reduced precision operations are accounted for during training.
Figure~\ref{fig:ROCAUC} (right) shows that with a QAT library called \brevitas~\citep{brevitas}, only 7 total bits are needed for full performance.
Details of the QAT procedure can be found in the Supplementary Material.

Figure~\ref{fig:ROCAUC} also shows the effect of graph truncation and zero-padding on the edge classification performance.
Graph zero-padding, in which null nodes and edges are appended to a graph with too few nodes or edges, does not usually affect the classification performance.
With the exception of a few rare cases, it is possible to pad a graph such that the null nodes and edges form a completely disconnected graph from the original graph.
In this case, no messages are passed between the original and null subgraphs, and results for the original subgraph are the same.
Graph truncation, on the other hand, has a twofold effect on the IN's performance.
The first effect is that truncated nodes and edges are no longer factored into the creation or passing of messages, which clearly affects the final output. 
The second effect is that truncated edges can no longer be classified by the IN.
To account for this second effect in our performance metrics, we consider each truncated edge as classified as false.
Figure~\ref{fig:ROCAUC} demonstrates that net result of the above effects is a drop in the optimal AUC from 99.9\% to about 95\%.

All resource estimates are computed using Vivado HLS 2019.2 using logic synthesis targeting a Xilinx Virtex UltraScale+ VU9P FPGA with part number \texttt{xcvu9p-flga2104-2L-e}.
The latency and II estimates are from C synthesis.
For simplicity, when scanning the total bit width \texttt{X} in the following results, we use a fixed-point precision of \apfixed{X}{X/2}, i.e. \texttt{X/2} bits are used for each the integer and fractional parts, but the results generalize to other quantization schemes.
The \hlsfml version used is a custom branch available at \cite{abdelabd_hls4ml}.

\smallskip
\subsection{Throughput-optimized results}

\begin{figure*}[!htbp]
\begin{center}
  \includegraphics[width=0.49\textwidth]{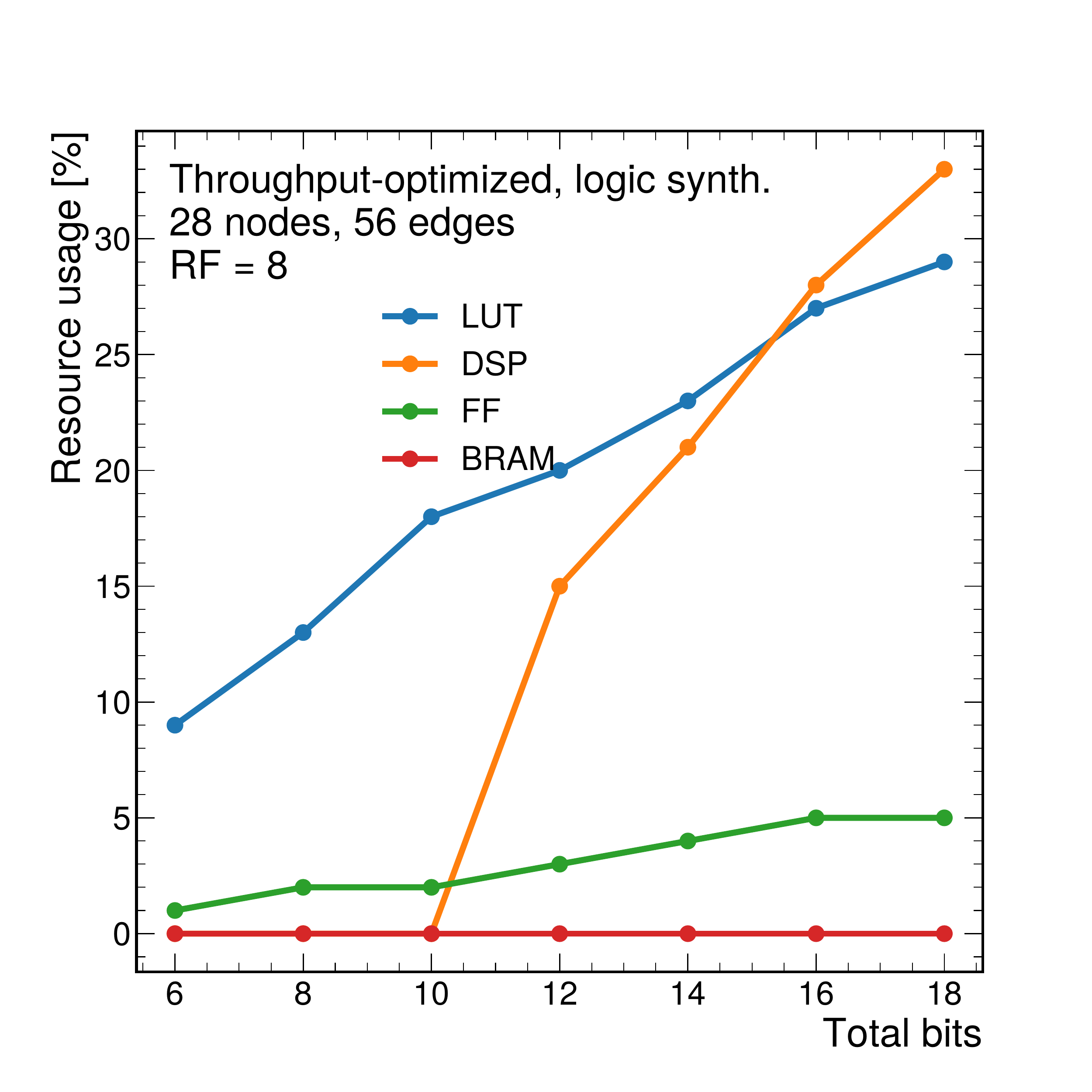}
  \includegraphics[width=0.49\textwidth]{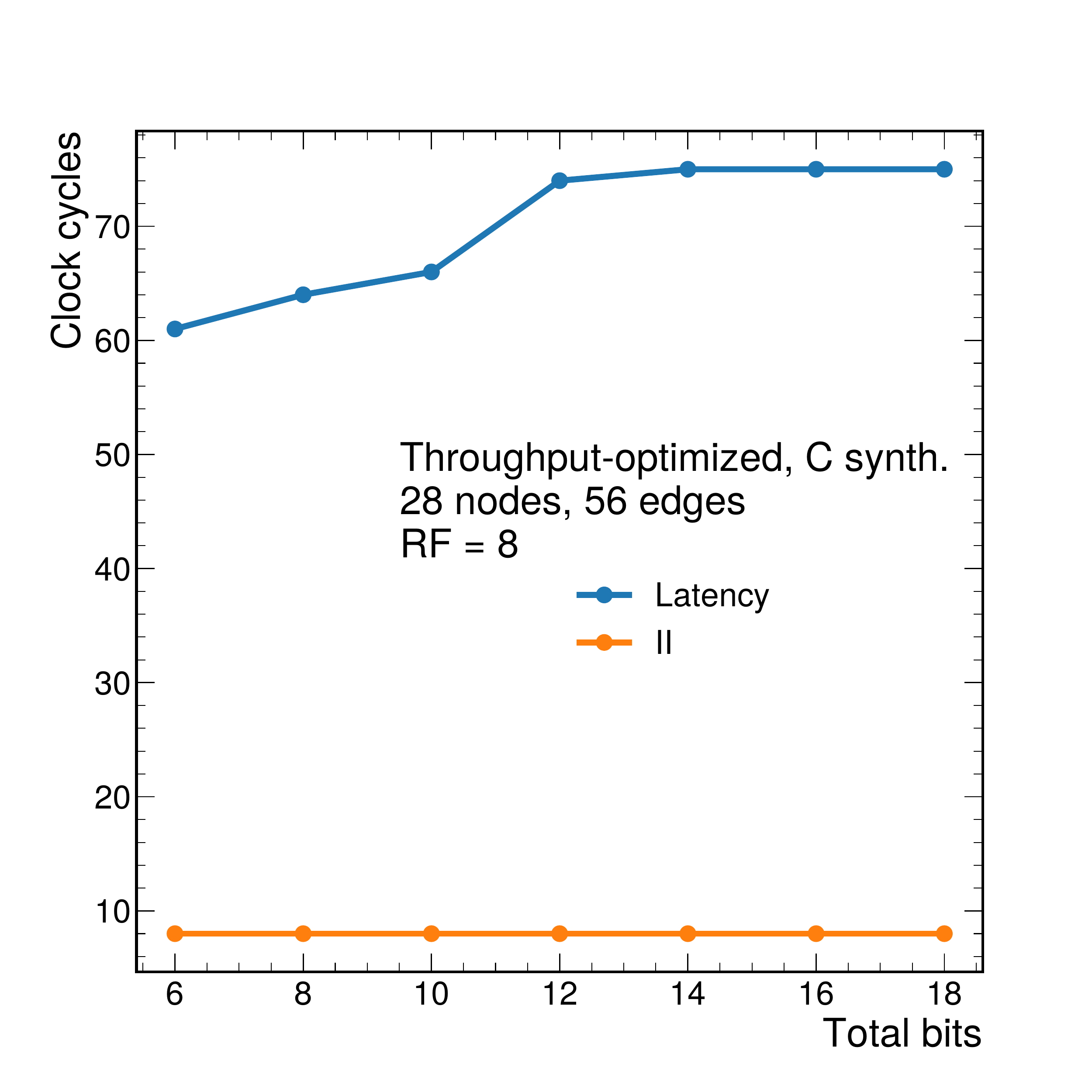}
  \end{center}
\caption{Resource usage estimates as a percentage after logic synthesis (left) and latency and II in clock cycles (right) for a constant reuse factor of 8 as a function of the total bit width.}
  \label{fig:throughput_precision_scan}
\end{figure*}

First, we present results for the throughput-optimized implementation.
We consider graphs consisting of 28 nodes and 56 edges, which is near the upper limit that can be synthesized with this design.
Figure~\ref{fig:throughput_precision_scan} (left) shows the resource usage as a function of the bit width for a constant RF of 8.
As expected, increasing the bit width increases the resource usage especially for LUTs and DSPs.
As shown previously, this model has good performance with a total bit width of 12 bits, however the bit width can be reduced down to 7 bits using QAT~\citep{Coelho:2020zfu,qkeras,Hawks:2021ruw}, meaning a substantion reduction in resource usage.
We also note that Vivado HLS implements multiplications with bit widths of 10 and above using DSPs, and multiplications below 10 bits using LUTs~\citep{vivado20192}.
For this reason, we see that the DSP usage drops to zero for 10 bits and below.

Figure~\ref{fig:throughput_precision_scan} (right) shows the latency in clock cycles (for a 5\unit{ns} clock period) as a function of the total bit precision, which ranges from about 300 to 370\unit{ns}.
For simplicity, we consider \apfixed{X}{X/2} data types, so the number of integer bits is half of the total bits.
By construction, the II for this design should be equal to the RF, although it may be smaller due to optimizations in Vivado HLS.
In this case, the II is constant at 40\unit{ns} given the constant RF of 8.

\begin{figure*}[!htbp]
\begin{center}
  \includegraphics[width=0.49\textwidth]{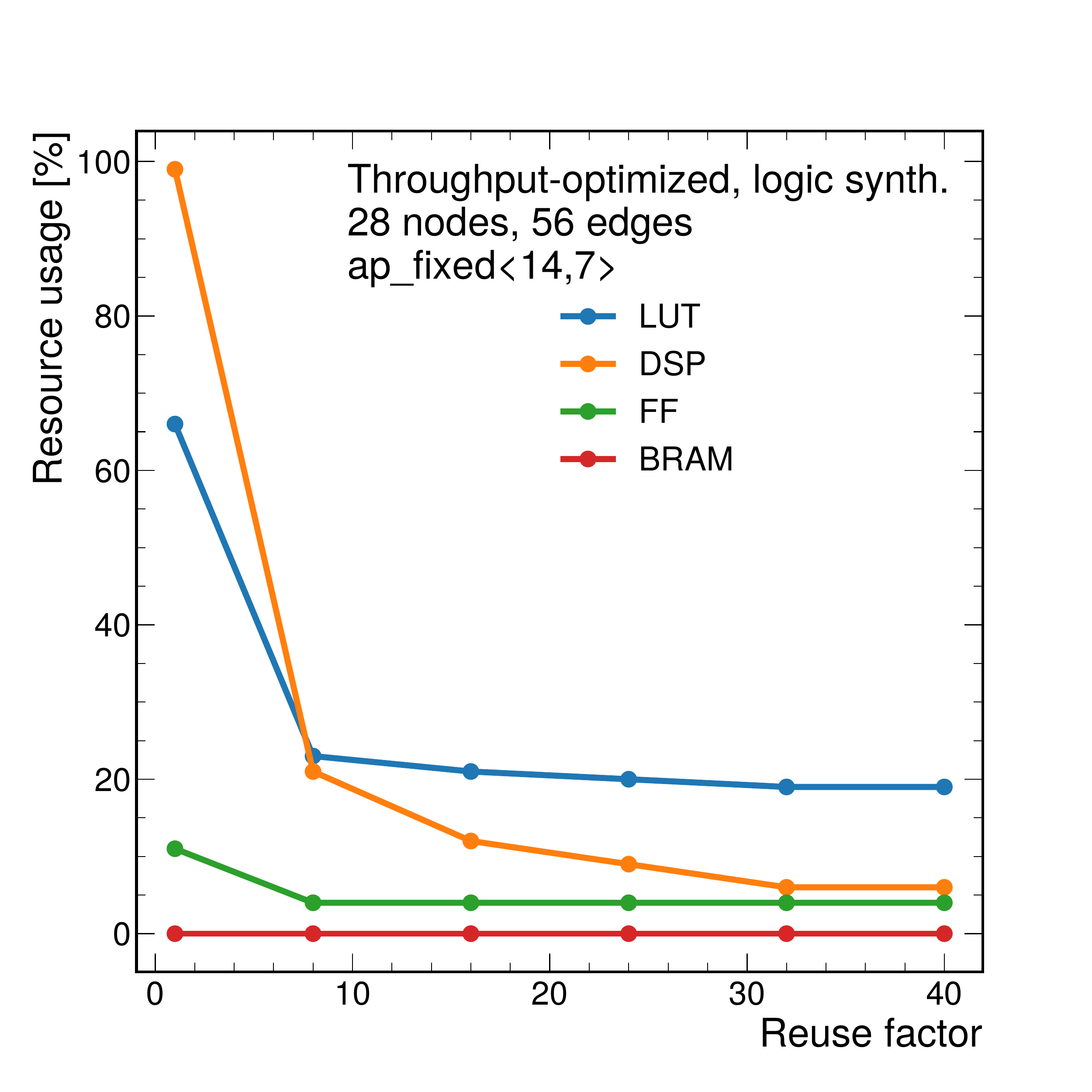}
  \includegraphics[width=0.49\textwidth]{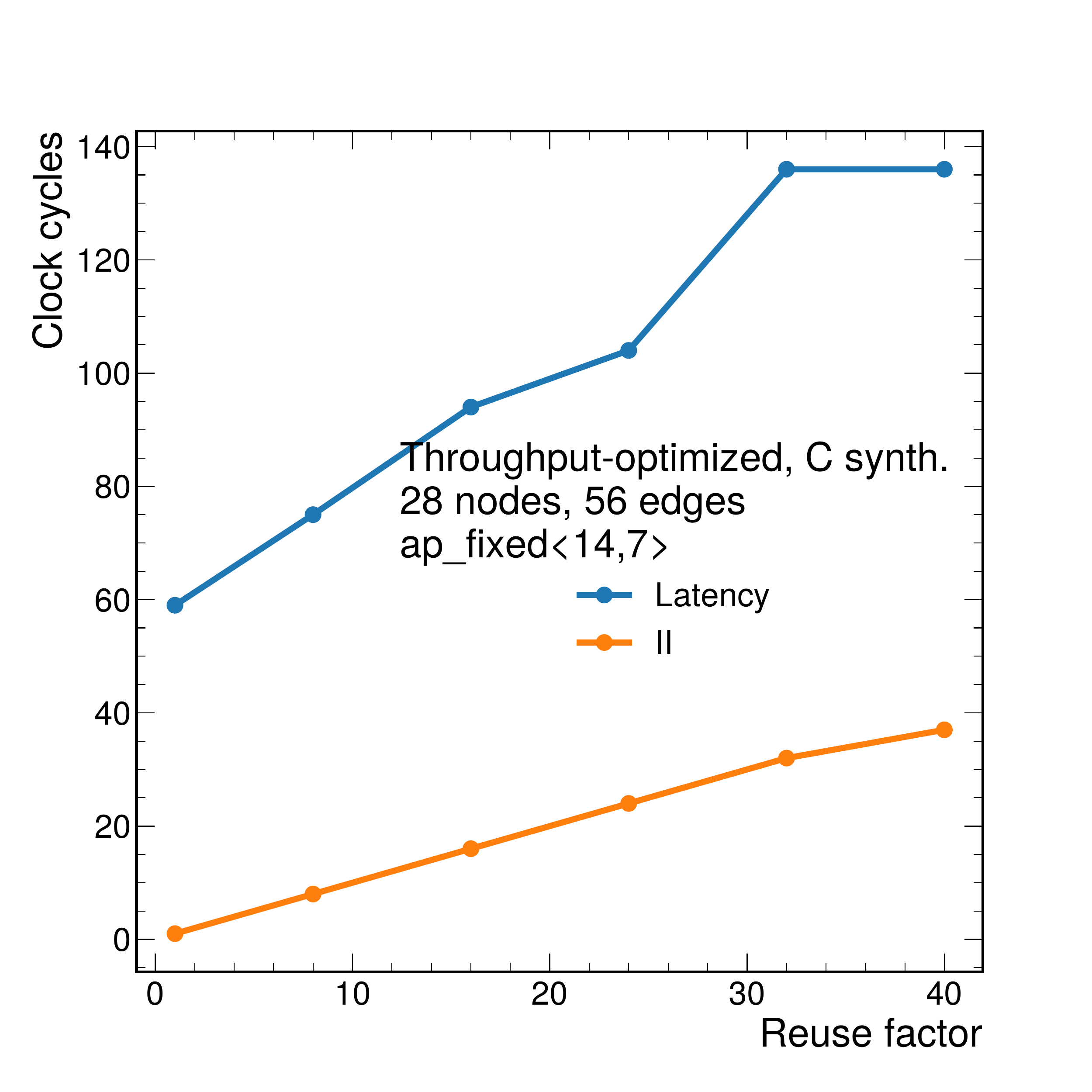}
  \end{center}
\caption{Resource usage estimates as a percentage after logic synthesis (left) and latency and II in clock cycles (right) for a constant fixed point precision of \apfixed{14}{7} as a function of the reuse factor.}
  \label{fig:throughput_reusefactor_scan}
\end{figure*}

We also scan the RF at a constant fixed point precision of \apfixed{14}{7}, to study the resources and timing as a function of decreasing concurrency.
Figure~\ref{fig:throughput_reusefactor_scan} shows the resource usage estimates (left) and latency and II (right) versus RF.
In general, increasing the RF, decreases the resources, while increasing the latency and II.
For a RF of 1, the algorithm saturates the FPGA (100\% DSP usage and 65\% LUT usage). However, increasing the reuse factor to 8, makes the algorithm much more feasible, with the same resources taking up about 25\% of the available ones.
As we scan the RF from 1 to 40, we find the latency ranges from about 400 to 700\unit{ns}, while the II ranges from 5 to 200\unit{ns}.

\begin{figure*}[!htbp]
\begin{center}
  \includegraphics[width=0.49\textwidth]{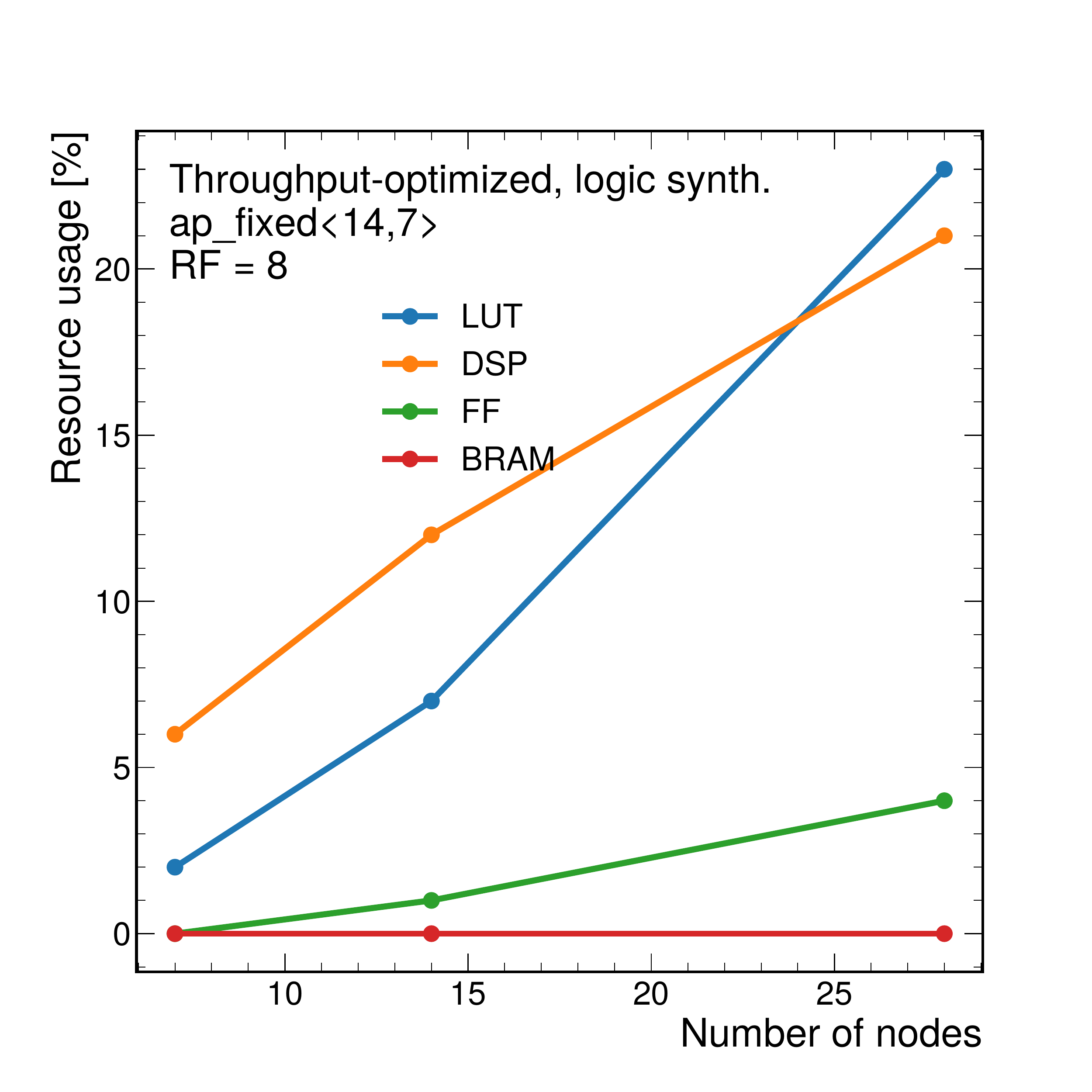}
  \includegraphics[width=0.49\textwidth]{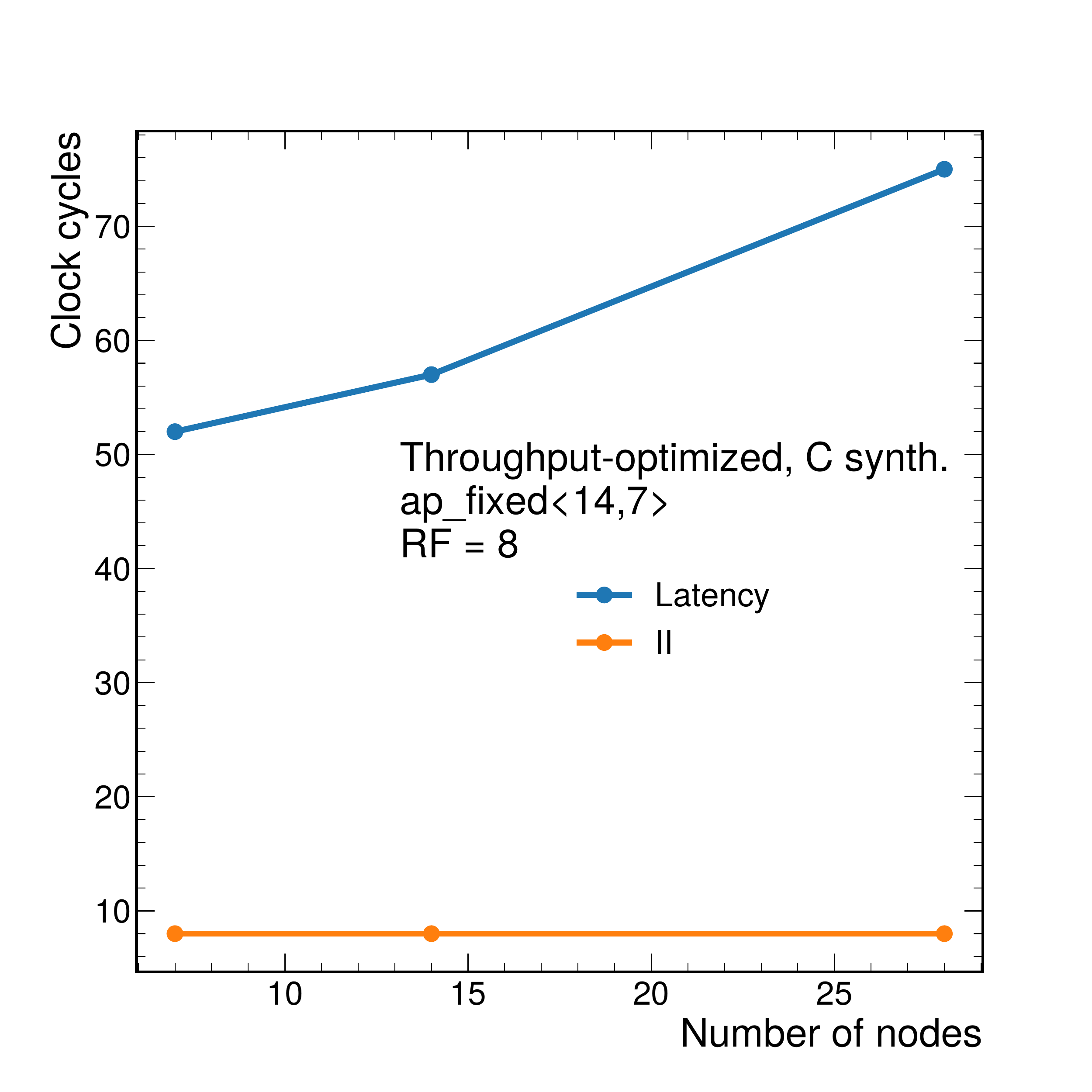}
  \end{center}
\caption{Resource usage estimates as a percentage after logic synthesis (left) and latency and II in clock cycles (right) for a constant reuse factor of 8 and a bit width of \apfixed{14}{7} as a function of the number of nodes $\nnodes$.
The number of edges is fixed to $\nedges = 2\nnodes$, as is empirically observed for the 2\GeV graphs.
Each clock cycle corresponds to 5\unit{ns}.}
  \label{fig:throughput_size_scan}
\end{figure*}

Figure.~\ref{fig:throughput_size_scan} also shows how the design scales as a function of the number of nodes $\nnodes$ from 7 to 28.
The number of edges $\nedges$ is fixed to $\nedges = 2\nnodes$, a relationship that is empirically observed for the 2\GeV graphs.
To synthesize larger graphs than those shown here, a different approach must be adopted. 

\subsection{Resource-optimized results}
\label{sec:resource_results}

\begin{figure*}[!htbp]
\begin{center}
  \includegraphics[width=0.49\textwidth]{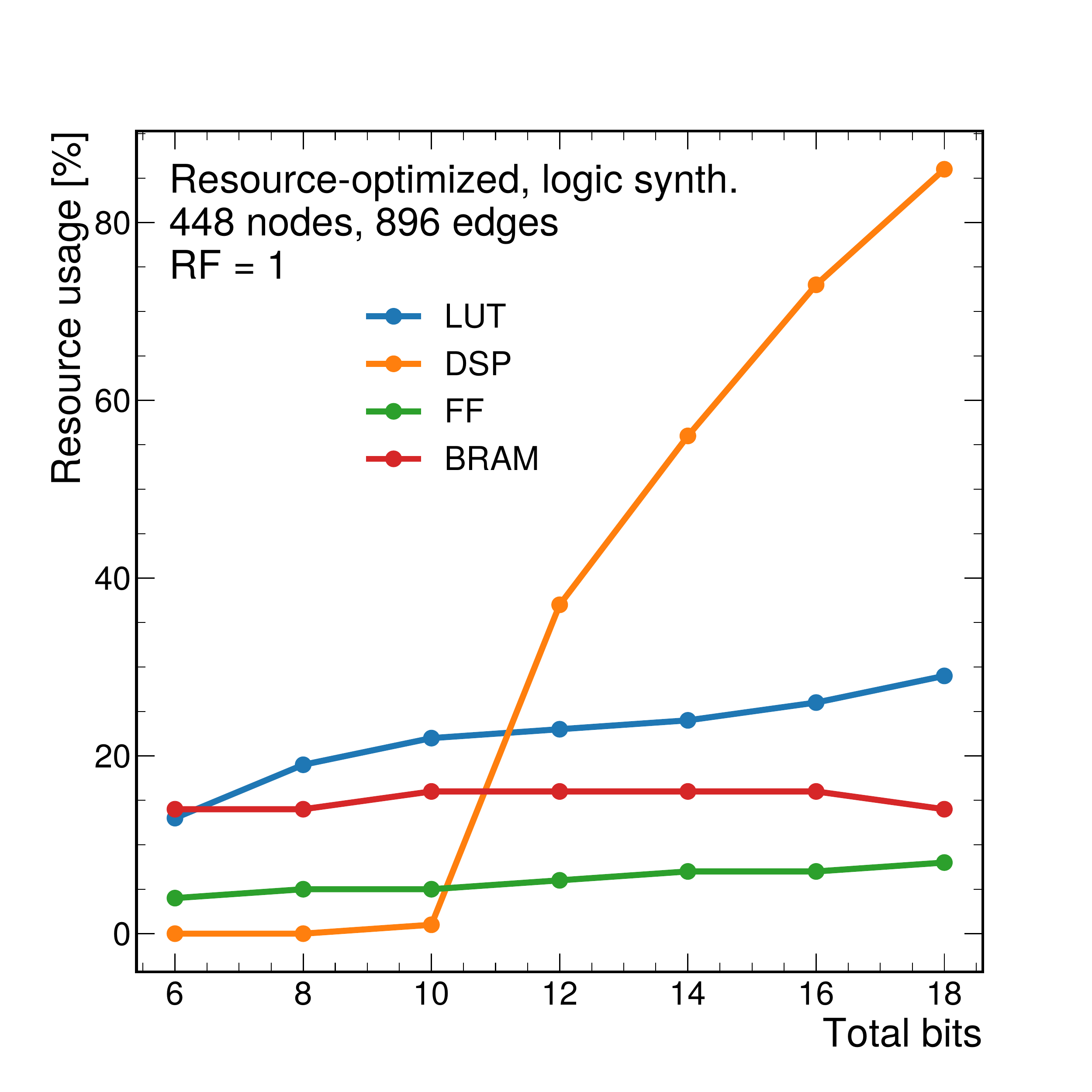}
  \includegraphics[width=0.49\textwidth]{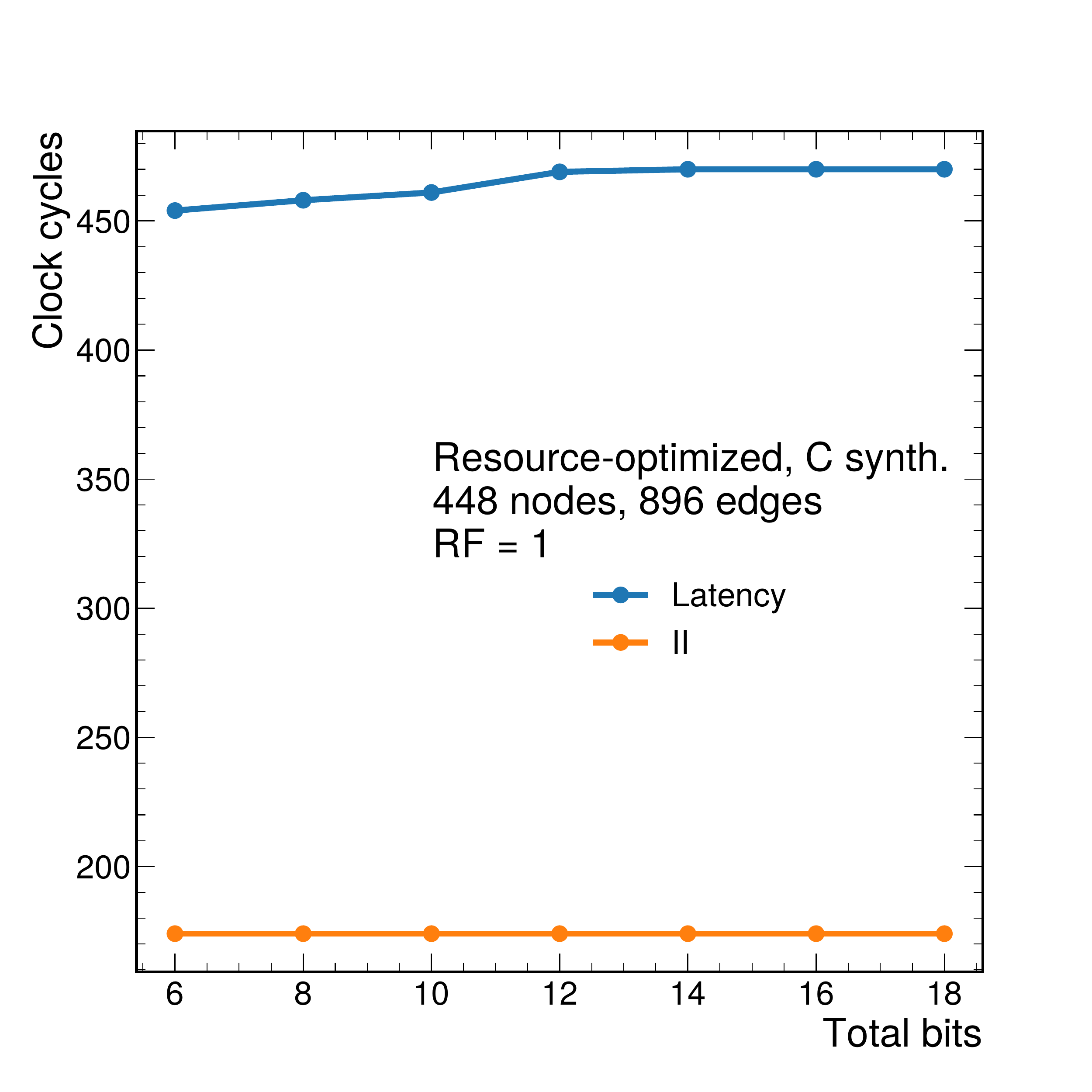}
  \end{center}
\caption{Resource usage estimates as a percentage after logic synthesis (left) and latency and II in clock cycles (right) for a constant reuse factor of 1 as a function of the total bit width for the resource-optimized implementation.
Input graphs consist of 448 nodes and 896 edges.
Each clock cycle corresponds to 5\unit{ns}.}
  \label{fig:resource_precision_scan}
\end{figure*}

\begin{figure*}[!htbp]
\begin{center}
  \includegraphics[width=0.49\textwidth]{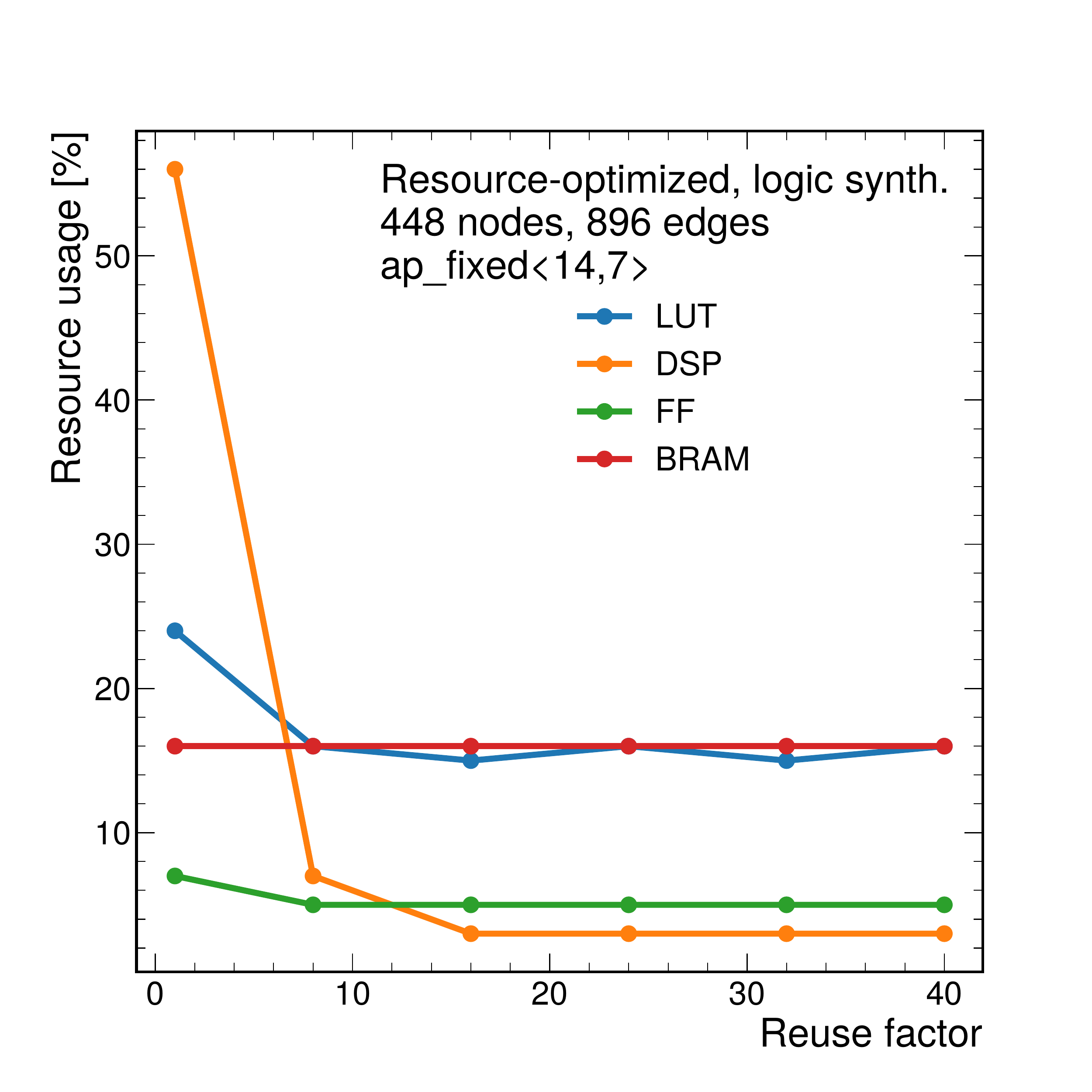}
  \includegraphics[width=0.49\textwidth]{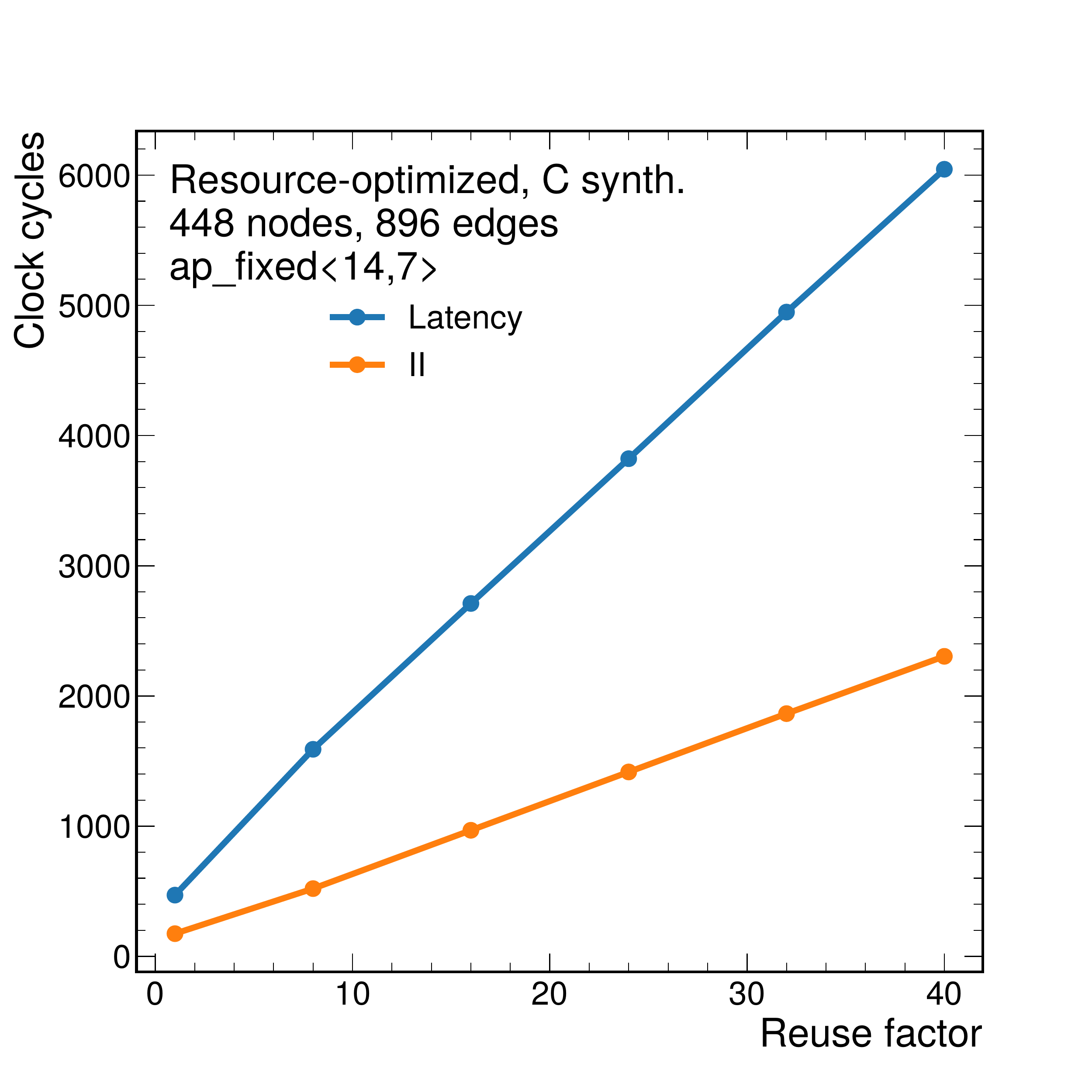}
  \end{center}
\caption{Resource usage estimates as a percentage after logic synthesis (left) and latency and II in clock cycles (right) for a constant fixed point precision of \apfixed{14}{7} as a function of the reuse factor for the resource-optimized implementation. 
Input graphs consist of 448 nodes and 896 edges.
Each clock cycle corresponds to 5\unit{ns}.}
  \label{fig:resource_reusefactor_scan}
\end{figure*}

To demonstrate how the design scales, results for the resource-optimized implementation are shown for 448 nodes and 896 edges are shown in Figures~\ref{fig:resource_precision_scan} and \ref{fig:resource_reusefactor_scan}.
For all results shown, we take the PF$=16$, but this is fully configurable by the user.
Despite the large graph size, the resources remain below 90\% for all bit widths and reuse factors considered.
However, the latency ranges from about 2.4\unit{$\mu$s} to 40\unit{$\mu$s} depending on the reuse factor.
Similarly the II ranges from about 850\unit{ns} to 11\unit{$\mu$s}.
While these latency and II results may be too long for L1 applications, they represent substantial improvements over CPU-based processing and thus may form the basis of a CPU-FPGA coprocessing workflow for a particle tracking application in the high-level trigger or offline processing.
For CPU-based inference of the same model (using a 12-core Intel Xeon CPU E5-2650 v4 @ 2.20\,GHz), the latency is about 1.06\unit{ms} per graph for the same size graphs (448 nodes and 896 edges) in the \texttt{PyG} framework~\citep{PyTorchGeometric}.

\begin{figure*}[!htbp]
\begin{center}
  \includegraphics[width=0.49\textwidth]{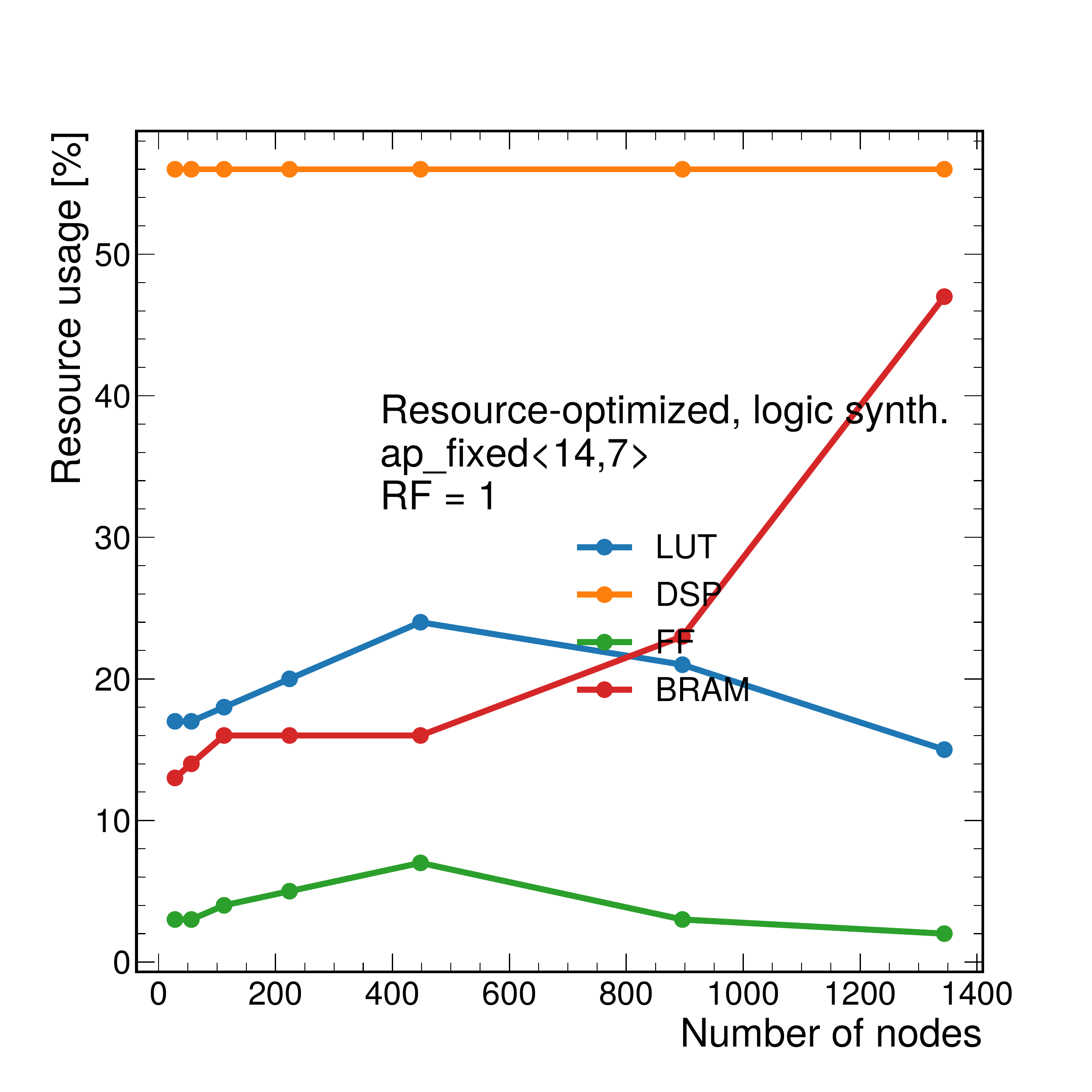}
  \includegraphics[width=0.49\textwidth]{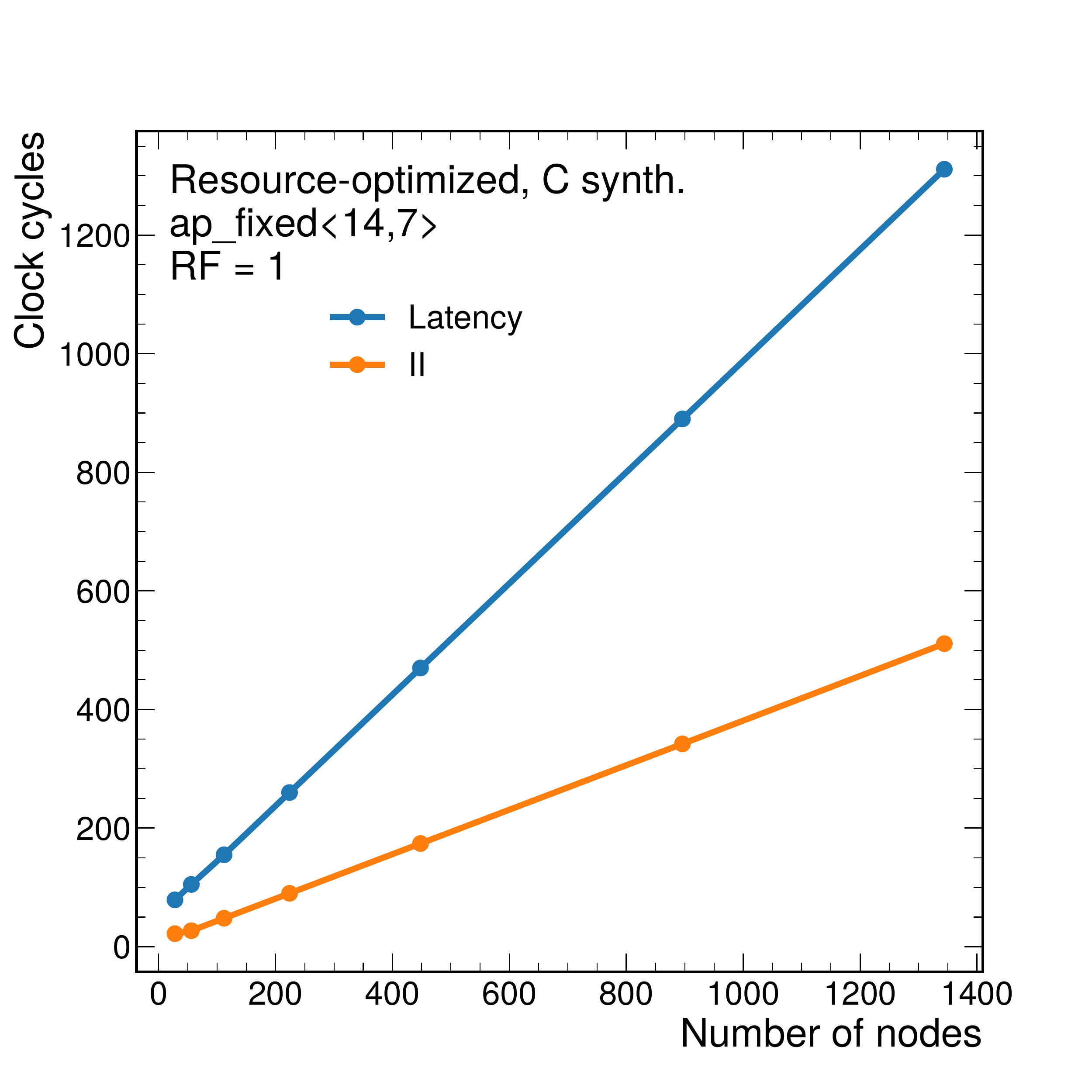}
  \end{center}
\caption{Resource usage estimates as a percentage after logic synthesis (left) and latency and II in clock cycles (right) for a constant reuse factor of 1 and a bit width of \apfixed{14}{7} as a function of the number of nodes $\nnodes$.
The number of edges is fixed to $\nedges = 2\nnodes$, as is empirically observed for the 2\GeV graphs.
Each clock cycle corresponds to 5\unit{ns}.}
  \label{fig:resource_size_scan}
\end{figure*}

Figure~\ref{fig:resource_size_scan} demonstrates the scalability of this resource-optimized design to larger graphs.
We vary the number of nodes from 28 to 1,344, while, as before, we fix $\nedges=2\nnodes$.
Given the dataflow design, the resources stay fairly consistent throughout, always staying below 60\% for the dominant resource (DSPs) when $RF$=1.
The latency and II scale linearly with the number of nodes.

Table~\ref{tab:synthesis_summary} summarizes the latency, II, and FPGA resource usage metrics of the synthesized firmware, in particular comparing the throughput-optimized design with the resource-optimized one for a few different configuration choices.
In particular, the table compares the throughput-optimized design for RF$=1$ and RF$=8$, noting the large reduction in resources (and relatively small increase in latency).
We also show how the resources remain stable for the resource-optimized design even when scaling to much larger graph sizes.
Finally, relative to the resource-optimized design, the throughput-optimized design is able to achieve the smallest latency and II for small graphs.

\begin{table}[hbtp]
    \caption{Summary of the latency, II, and FPGA resource usage metrics of the synthesized firmware for a variety of design choices.
    The target FPGA is a Xilinx Virtex UltraScale+ VU9P FPGA (part number \texttt{xcvu9p-flga2104-2L-e}), which has 6,840 DSPs, 1,182,240 LUTs, 2,364,480 FFs, and 75.9\unit{Mb} of BRAM~\citep{datasheet}.
    A 5\unit{ns} clock period is used.}
    \label{tab:synthesis_summary}
    \begin{center}
    \resizebox{\textwidth}{!}{
    \begin{tabular}{lccc|cccccc}
    \hline
    \multirow{2}{*}{Design} & \multirow{2}{*}{(\nnodes, \nedges)} & \multirow{2}{*}{RF} & \multirow{2}{*}{Precision} & Latency & II & \multirow{2}{*}{DSP [\%]} & \multirow{2}{*}{LUT [\%]} & \multirow{2}{*}{FF [\%]} & \multirow{2}{*}{BRAM [\%]} \\
    & & & & [cycles] & [cycles] & & & & \\
    \hline
    Throughput-opt. & (28, 56) & 1 & \apfixed{14}{7} & 59 & 1 & 99.9 & 66.0 & 11.7 & 0.7 \\
    Throughput-opt. & (28, 56) & 8 & \apfixed{14}{7} & 75 & 8 & 21.9 & 23.8 & 4.7 & 0.7 \\
    Resource-opt. & (28, 56) & 1 & \apfixed{14}{7} & 79 & 28 & 56.6 & 17.6 & 3.9 & 13.1\\
    Resource-opt. & (448, 896) & 1 & \apfixed{14}{7} & 470 & 174 & 56.6 & 25.0 & 7.4 & 16.5\\
    Resource-opt. & (448, 896) & 8 & \apfixed{14}{7} & 1590 & 520 & 5.6 & 25.0 & 7.4 & 16.3\\\hline
    \end{tabular}}
    \end{center}
\end{table}

\section{Summary and outlook}
\label{sec:summary}

In summary, we develop two complementary field-programmable gate array (FPGA) implementations of a graph neural network (GNN) for charged particle tracking at the LHC.
Namely, the GNN classifies track segments as true or false based on a graph constructed from the positions of hits in the tracking detector.
One of the implementations is optimized for low-latency and high-throughput typical for applications in the FPGA-based level-1 trigger systems at the LHC.
The other implementation is optimized to minimize the FPGA resources needed and is capable of scaling to much larger graph sizes (thousands of nodes and edges).
In order to make this possible, multiple improvements were made including an optimization of the memory access of the input data and the instantiation of multiple parallel processing engines. 
This implementation is applicable for FPGA-CPU coprocessing workflows in both the software-based high-level trigger and offline computing.
The conversion of the trained model, specified using \textsc{PyTorch Geometric}, into high-level synthesis (HLS) code is achieved automatically using a custom converter integrated into \hlsfml, a source-to-source compiler.

Several further improvements can reduce the resource usage or improve the performance.
In particular, we showed quantization-aware training (QAT)~\citep{Coelho:2020zfu,qkeras,Hawks:2021ruw} can significantly reduce the number of bits required, but further work is needed to fully support QAT GNN models into \hlsfml.
A further detector-motivated optimization is the restriction of which nodes can be accessed simultaneously.
Currently, the firmware is generic enough to allow any two nodes in the input graph to be connected, but given the detector geometry and hitgraph construction, certain nodes will never be directly connected (such as those belonging to the same tracker layer or on opposite sides of the detector).
Implementing this restriction should further reduce the FPGA resources required.

Other considerations are necessary for implementing such a trigger in real experimental settings. 
For instance, graph construction, track building, and track fitting are additional steps that need to be applied in addition to the track segment classification performed here.
Further, the sectors we use are non-overlapping, which means boundary effects can reduce the accuracy of the track segment classification.
To resolve this, a typical technique is to use overlapping regions and define ``fiducial regions'' as the central portions of each region, where boundary effects are less significant.
Further duplicate removal may be necessary as well when building full track candidates.

Despite these caveats and further possible improvements, this study demonstrates that for small graphs with dozens of nodes and edges, inference of GNNs for charged particle tracking is possible within the strict sub-microsecond latency and FPGA resource requirements of the level-1 trigger at the LHC.
Further, for CPU-FPGA coprocessing applications, larger graphs with thousands of nodes and edges can also be processed with latencies in the tens of microseconds range, which still represent a considerable speedup with respect to CPU-only inference.

\section*{Data Availability Statement}
Publicly available datasets were analyzed in this study. 
This data can be found here: \href{https://kaggle.com/c/trackml-particle-identification}{https://kaggle.com/c/trackml-particle-identification}.

\section*{Author Contributions}
AE and VR developed the \hlsfml implementation of the neural networks, especially the throughput-optimized design, with input and supervision from JD, MN, MA, ST, GD, IO, and PE.
S-YH improved and developed the resource-optimized HLS implementation with input and supervision from B-CL, J-XH, and S-CH.
MT developed and trained the quantization-aware models with input and supervision from S-YH, SH, and JD.
All authors contributed to the article and approved the submitted version.

\section*{Funding}
This work was supported by IRIS-HEP through the U.S. National Science Foundation (NSF) under Cooperative Agreement OAC-1836650.
JD was supported by the U.S. Department of Energy (DOE), Office of Science, Office of High Energy Physics Early Career Research program under Award No. DE-SC0021187.
GD was supported by DOE Award No. DE‐SC0007968. 
B-CL was supported by the Taiwan Ministry of Science and Technology under MOST 110-2224-E-A49-004.
S-CH and SH were supported by NSF Award No. OAC-1934360.
MA and MN were supported by NSF Award No. OAC-1934757.

\section*{Acknowledgments}
We gratefully acknowledge the input and discussion from the Exa.TrkX collaboration.
We also acknowledge the Fast Machine Learning collective as an open community of multi-domain experts and collaborators. 
This community was important for the development of this project.
In particular, Vladimir Loncar, Sioni Summers, Duc Hoang, Yutaro Iiyama, Maurizio Pierini, Nhan Tran, Philip Harris, Mia Liu, Sofia Vallecorsa, and Kazi Ahmed Asif Fuad provided valuable input for the \hlsfml implementation of the graph neural network.

\bibliographystyle{frontiersSCNS} 
\bibliography{references}

\noindent\textbf{Conflict of Interest}: The authors declare that the research was conducted in the absence of any commercial or financial relationships that could be construed as a potential conflict of interest.
\ifarxiv
\clearpage
\setcounter{section}{0}
\setcounter{figure}{0}
\renewcommand{\thefigure}{S\arabic{figure}}
{\huge\helveticaitalic{\textbf{Supplementary Material}}}

\section{Alternative graph construction}
\label{sup:pt1GeV}

In this supplementary material, we investigate an alternative graph construction to show how the algorithm may be applied in different particle tracking settings, specifically for $\pt>1\GeV$ graphs.
The graph construction restrictions are $z_0 < 350$\unit{mm} and $\Delta\phi / \Delta r < 0.0007$ and $\pt^{\mathrm{min}} = 1\GeV$.
Each event is divided into 8 $\phi$ sectors, and 8 $\eta$ sectors.
Example graphs for 2 sectors in one event are shown in Fig.~\ref{fig:example_graphs_1GeV}.

\begin{figure*}[!htpb]
    \begin{center}
    \includegraphics[width=0.49\textwidth]{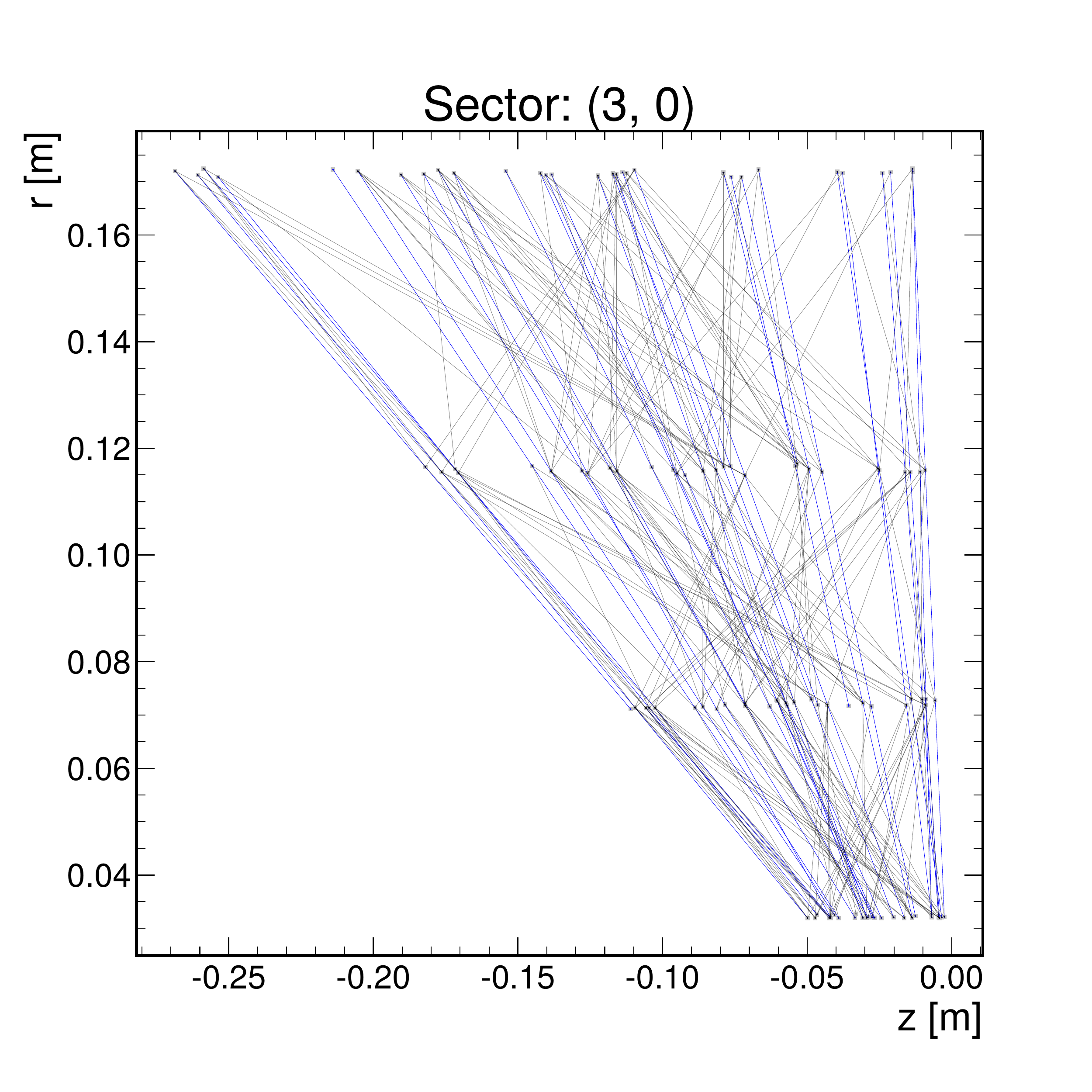}
    \includegraphics[width=0.49\textwidth]{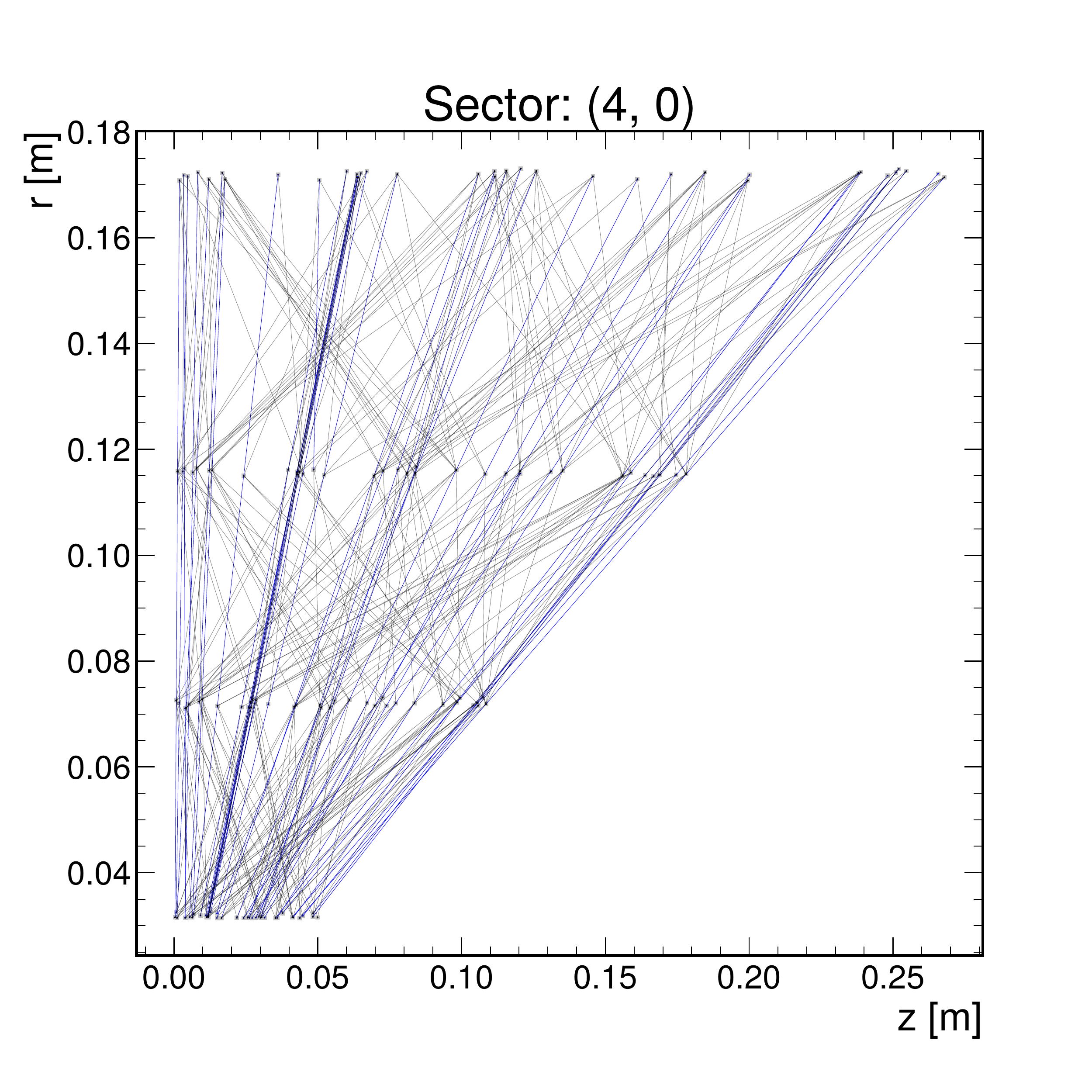}
    \end{center}
    \caption{Example graphs showing 2 of the sectors for one event with $\pt^\mathrm{min} = 1\GeV$, $z_0 < 350$\unit{mm}, $\Delta\phi / \Delta r < 0.0007$, 8 $\phi$ sectors, and 8 $\eta$ sectors.
    True track segments are denoted by blue edges, while false track segments are denoted by gray.}
    \label{fig:example_graphs_1GeV}
\end{figure*}

\begin{figure*}[!htpb]
    \begin{center}
    \includegraphics[width=0.49\textwidth]{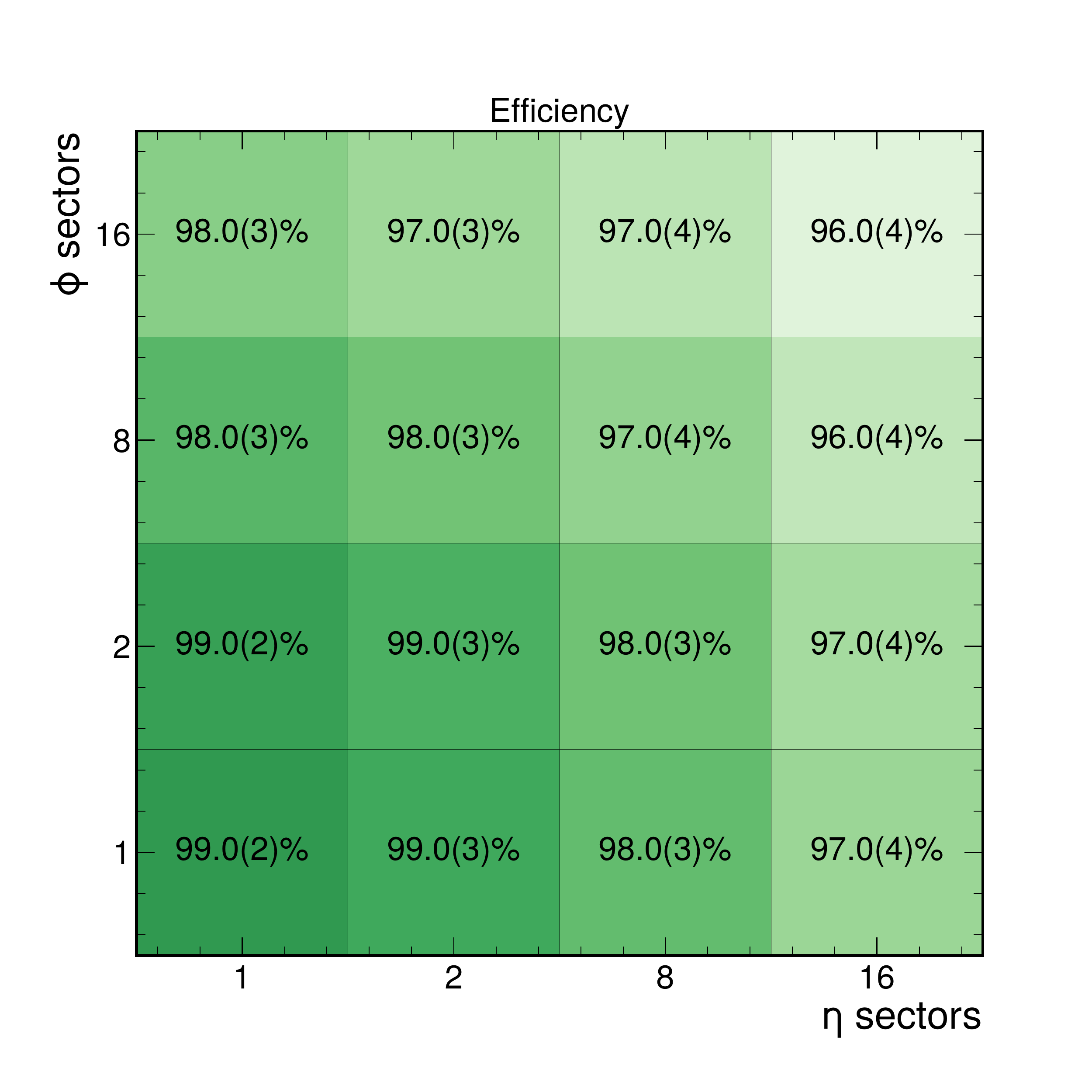}
    \includegraphics[width=0.49\textwidth]{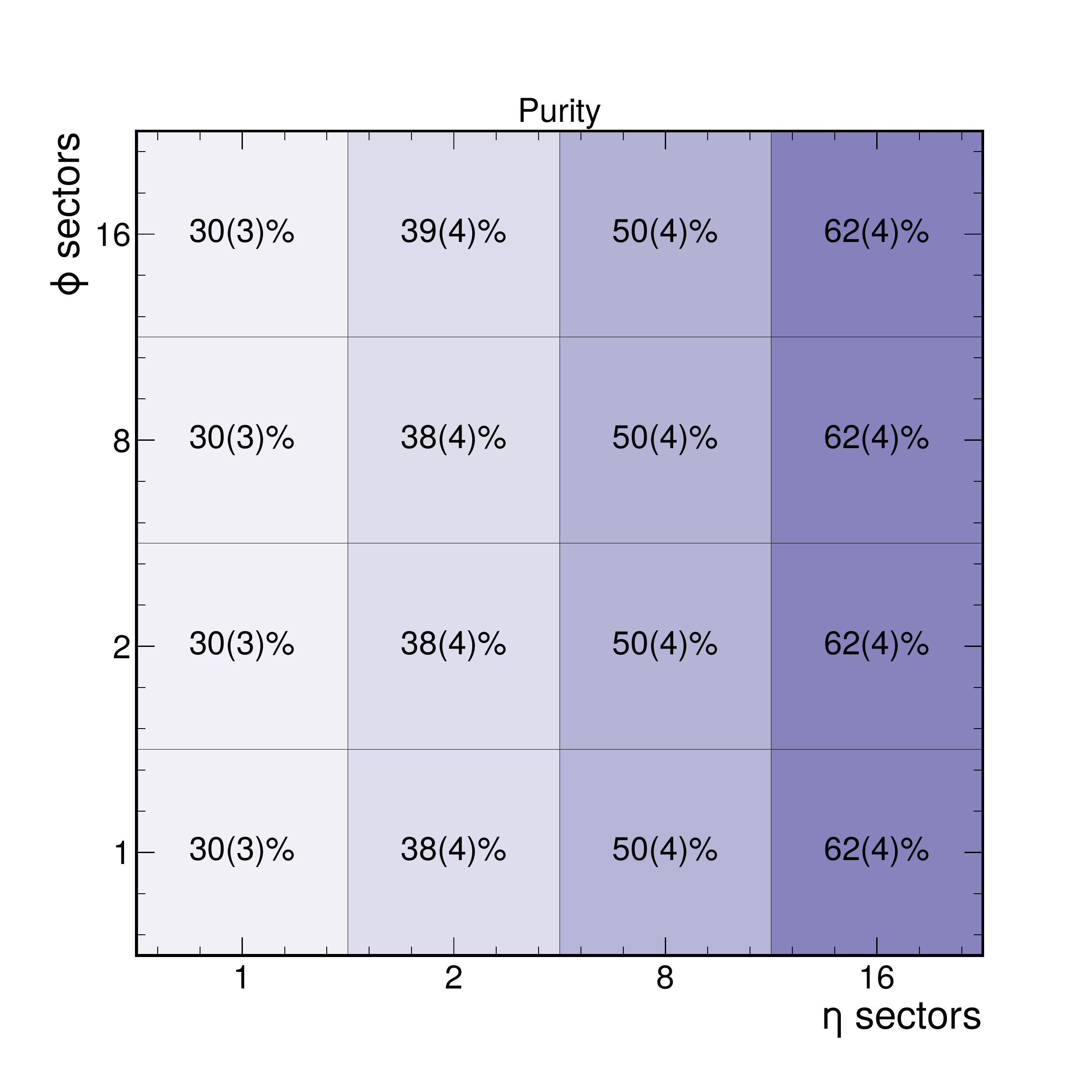}
    \end{center}
    \caption{
    Efficiency (left) and purity (purity) of the 1\GeV hitgraphs studied for different numbers of $\eta$ and $\phi$ sectors based on 50 events in \texttt{train\_1}.}
    \label{fig:efficiency_purity_1GeV}
\end{figure*}

The efficiency and purity of the graph construction method are shown for different choices for the number of $\phi$ and $\eta$ sectors in Fig.~\ref{fig:efficiency_purity_1GeV} based on 50 events in \texttt{train\_1}.
In particular for 8 $phi$ sectors and 8 $\eta$ sectors, the graphs retain an efficiency of 97\% and a purity of 50\%.

Figure~\ref{fig:graphs_1GeV} shows the 95th percentile for the number of nodes and edges in each sector depending on the number of sectors chosen.
For example, the average 95th percentile graph size for 8 $\phi$ sectors and 8 $\eta$ sectors is 162 nodes and 326 edges for this graph construction.
However, we note the distribution of nodes and edges depends on the particular $|\eta|$ range of the sector.

\begin{figure*}[!htbp]
\begin{center}
  \includegraphics[width=0.45\textwidth]{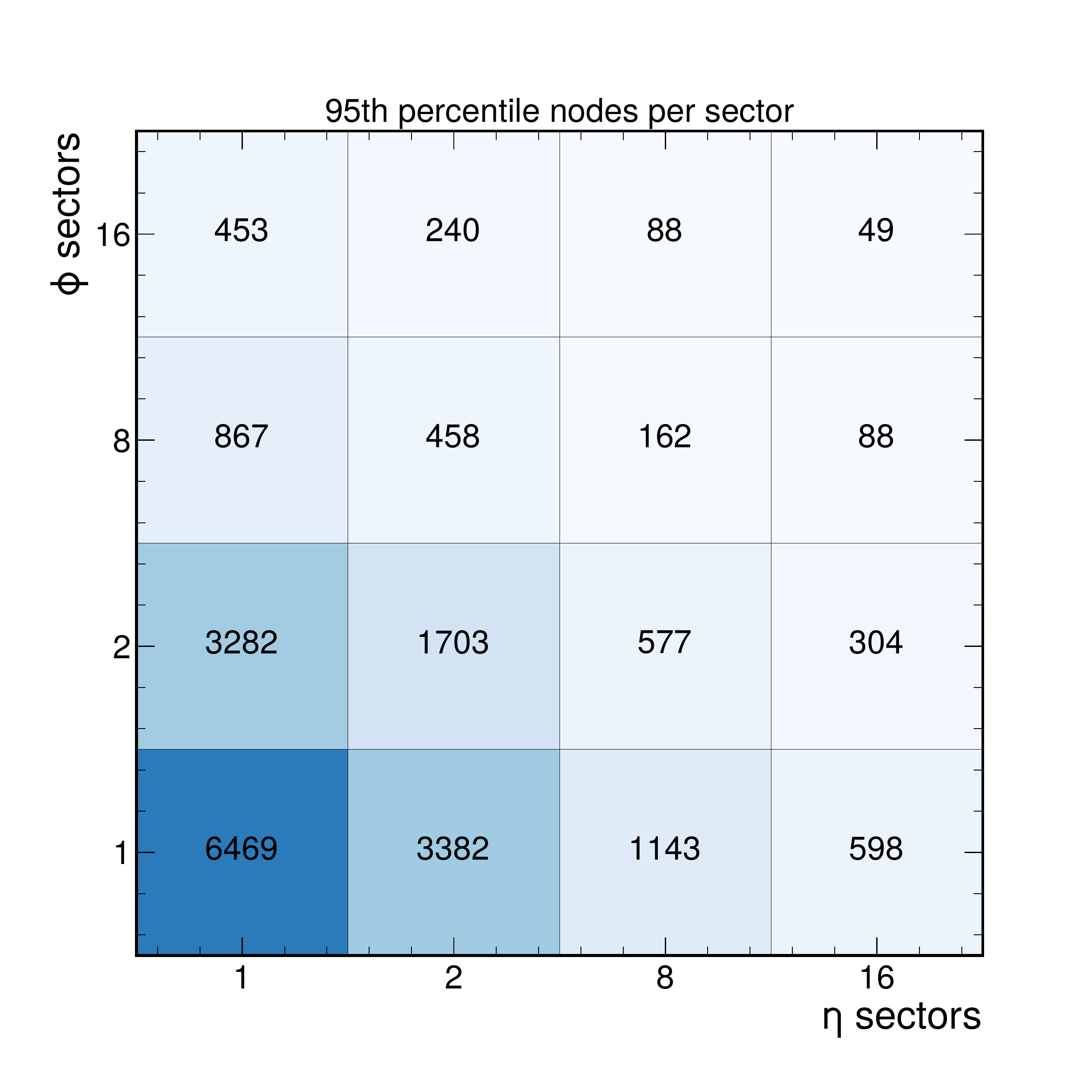}
  \includegraphics[width=0.45\textwidth]{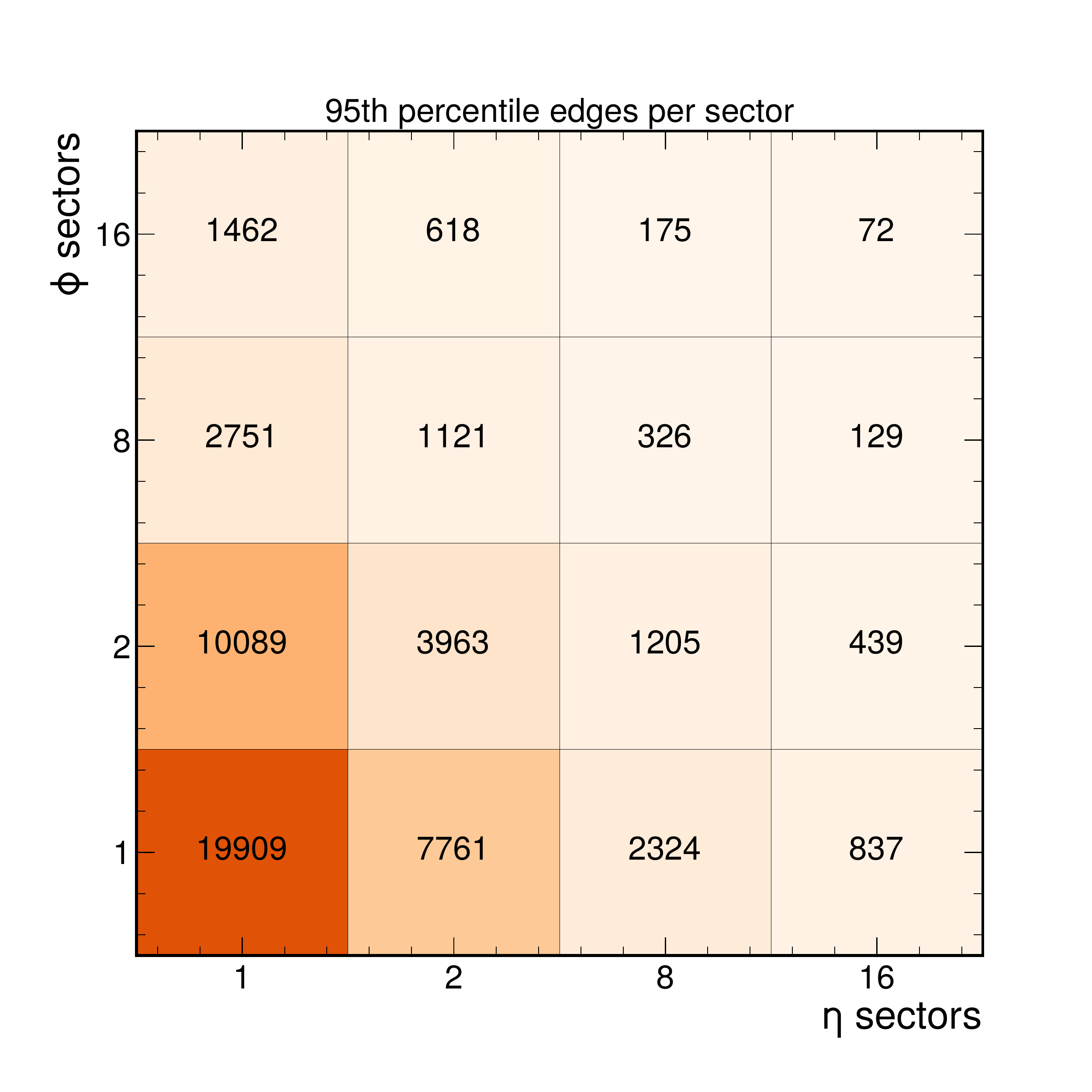}
  \end{center}
\caption{95th percentile of the number of nodes and edges in each sector for the 1\GeV graphs as a function of the number of $\eta$ and $\phi$ sectors, based on 50 events in \texttt{train\_1}.}
  \label{fig:graphs_1GeV}
\end{figure*}

Figure~\ref{fig:ROCAUC_1GeV} shows the AUC values for the 1\GeV graphs as a function of the total bit precision, where half of the available bits are used for the integer part and the other half are used for the fractional part.
Different from the 2\GeV task, we see that with 16 total bits, we reproduce the 32-bit floating point model when applying the \apfixed{X}{X/2} PTQ scheme.

\begin{figure*}[!htbp]
\begin{center}
  \includegraphics[width=0.45\textwidth]{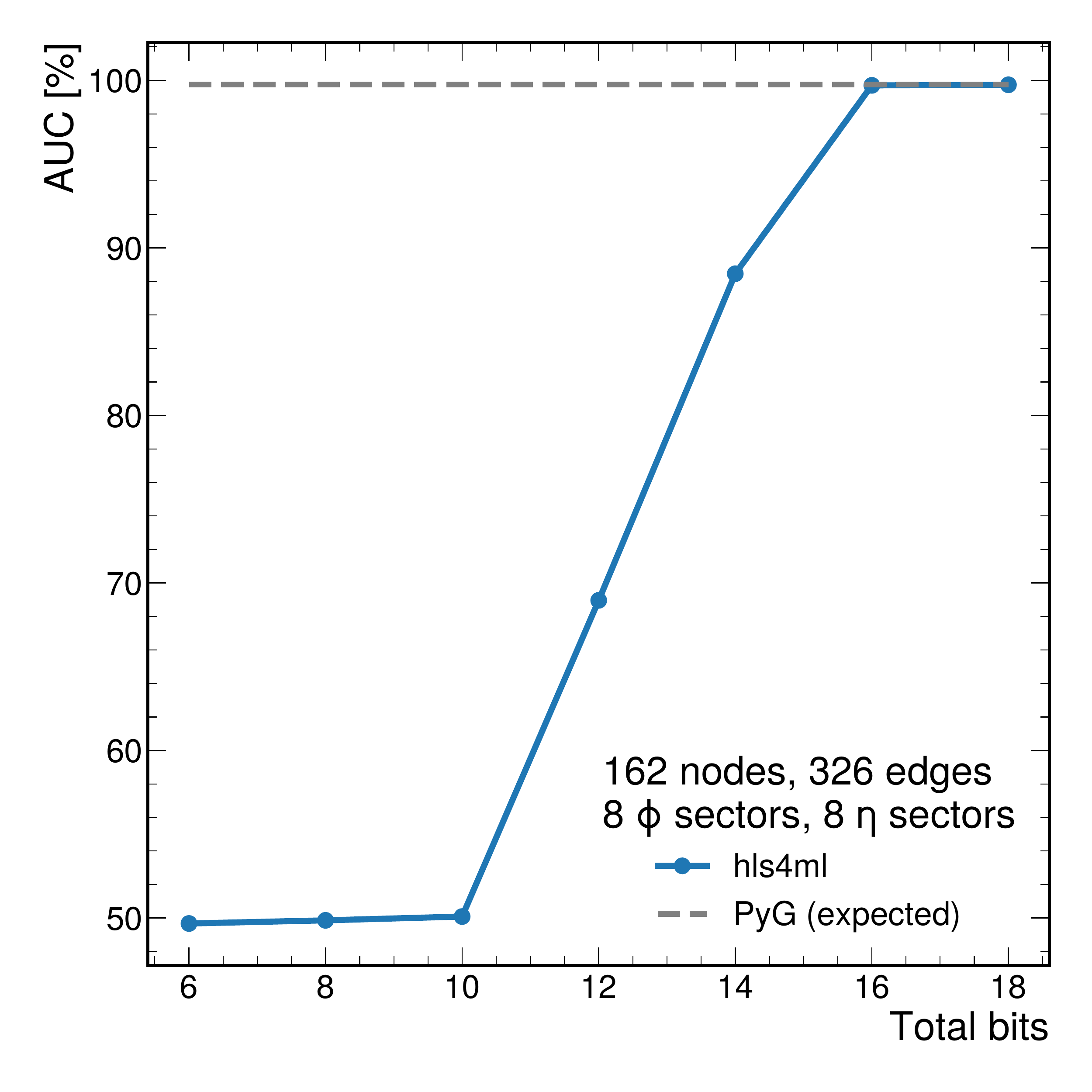}
  \end{center}
\caption{AUC values as a function of the total bit width \texttt{X} when using \apfixed{X}{X/2} with sectorized 1\GeV input graphs truncated at 162 nodes, 326 edges, corresponding to the 95\% percentile graph size.
The performance is evaluated with 1000 graphs from \texttt{train\_2}.
With precision greater than \apfixed{16}{8}, the AUC closely approximates the full floating point model.}
  \label{fig:ROCAUC_1GeV}
\end{figure*}

\section{Quantization-aware training}

Additional interaction network models were trained at different bit widths using the \brevitas library to illustrate benefits of QAT. 
\brevitas uses a scaled integer quantization scheme compared to the fixed-point precision scheme of \hlsfml, however we expect the hardware resources and timing to be comparable for the same number of total bits.
\brevitas implements scaled integer quantization by assigning a zero point and scale factor for all of the inputs and activations.
From there, the inputs and activations can be shifted and scaled to fit within the integer range defined by the bit width of the quantization.
In the case that an input or activation exceeds the minimum or maximum value of the scaled integer range, the value is clamped at the boundary.
As illustrated in the main text, the network retains the full performance even down to 7 total bits.
PTQ will inherently reduce accuracy due to the loss of information that occurs when converting from the floating-point representation to the fixed-point or scaled integer representation.
QAT allows for optimization while taking the loss of precision into account, allowing the network to train around the loss in precision and maintain accuracy.
For this study, QAT models with bit widths from 2 to 18 were trained using the same training set (\texttt{train\_1}) and evaluated on the same testing set (\texttt{train\_2}) as the PTQ models to generate ROC curves and calculate the AUC at different bit widths.
\fi
\end{document}